\documentclass[aps,twocolumn,groupedaddress,superscriptaddress]{revtex4-1}

\usepackage{graphicx,amsmath,amssymb}
\usepackage{soul}
\begin{document}
\begin{titlepage}
\title{Density dependence of relaxation dynamics in glass formers, and the dependence of their fragility on the softness of inter-particle interactions}

\author{Anshul D. S. Parmar}
\affiliation{Jawaharlal Nehru Center for Advanced Scientific Research, Jakkur Campus, Bengaluru 560064, India.}
\affiliation{TIFR Center for Interdisciplinary Sciences, 21 Brundavan Colony, Narsingi, Hyderabad  500075, India.}
\author{Pallabi Kundu}
\affiliation{Indian School of Mines (Indian Institute of Technology) Dhanbad, Jharkhand 826004, India.}
\author{Srikanth Sastry$^{*}$}
\affiliation{Jawaharlal Nehru Center for Advanced Scientific Research, Jakkur Campus, Bengaluru 560064, India.}

\begin{abstract}{
Fragility, quantifying the rapidity of variation of relaxation times, is analysed for a series of  model glass formers, which differ in the softness of their  interparticle interactions. In an attempt to rationalize experimental observations in colloidal suspensions that softer interactions lead to stronger (less fragile) glass formers, we study the variation of relaxation dynamics with density, rather than temperature, as a control parameter. We employ density-temperature scaling, analyzed in recent studies, to address the question. We find that while employing inverse density in place of temperature leads to the conclusion that softer interactions lead to stronger behaviour, the use of scaled variables involving temperature and density lead to the opposite conclusion, similarly to earlier investigations where temperature variation of relaxation dynamics was analysed for the same systems. We rationalize our results by considering the Adam-Gibbs (AG) fragility, which incorporates the density dependence of the configurational entropy and an activation energy that may arise from other properties of a glass former. Within the framework of the Adam-Gibbs relation, by employing density temperature scaling for the analysis, we find that softer particles make more fragile glasses, as deduced from dynamical quantities, which is found to be consistent with the  Adam-Gibbs fragility. }
\end{abstract} 
 \maketitle
\end{titlepage}

\section{Introduction}
Understanding the variation of quantities describing relaxation dynamics, the viscosity, various relaxation times, and diffusion coefficients, with control variables such as (most commonly) temperature, pressure or density, is central to the study of glass forming liquid and related glassy systems. It is well known that upon approaching the glass transition, these quantities show rapid variation, in excess of what may be expected, {\it e. g.} by the temperature dependence of time scales described by the Arrhenius law. The rapidity of such variation has been attempted to be captured by {\it fragility} \cite{fragility_angell} and has been the subject of considerable investigation in the last two decades \cite{fragility_symposium}.  Fragility has been quantified in various ways based on dynamical data 
\cite{fragility_angell,ruocco}, but a straight forward one to consider is through the VFT (Vogel-Fulcher-Tammann) relation that has been used to describe the behaviour of viscosity and other quantities. In order to define such quantification of fragility, termed {\it kinetic fragility},  the VFT relation can be written in the form 

\begin{equation}
B(T) = B_{0} \exp\left[\frac{1}{K^{(B)}_{VFT} \left (\frac{T}{T_{VFT}}-1 \right)} \right] 
\label{eqn:kinfr2}
\end{equation}
where $B$ is a quantity such as the viscosity.   
A description of the temperature dependence of $B$ as written  defines the {kinetic fragility $K^{(B)}_{VFT}$
($T_{VFT}$ is the VFT divergence temperature, which we assume is finite and does not depend on the dynamical quantity studied  
, {although fit values do. But}
 the definition of fragility does not critically depend on the presence of a divergence temperature). 

Analysis of fragility in terms of thermodynamic behaviour has also been performed ({\it e. g.} \cite{pap:AG-sastry,Martinez}) and relies on the Adam-Gibbs relation that expresses dynamical data in terms of the configurational entropy: 

\begin{equation}
B(T) = B_{0} \exp\left[\frac{A^{(B)}}{TS_{c}}\right].
\label{eqn:AG}
\end{equation}
The configurational entropy $S_c$ is the difference between the total entropy of a liquid and the``vibrational" entropy of individual ``glasses" or ``basins", which in what follows will be considered to be basins of local energy minima or {\it inherent structures} \cite{pap:AG-sastry}. Thus,  

\begin{equation}
S_c(T) = S_{total}(T) -S_{vib}(T).
\label{confent}
\end{equation}

If one assumes that the temperature dependence of $S_c$ is well described by 
\begin{equation} \label{eq:kagPEL}
T S_c(T) = K_{T}~\left ({T \over T_K} -1\right), 
\end{equation}
the Adam-Gibbs relation leads to the VFT relation, with the identification $T_{VFT} = T_K$, and $K_{VFT} = K_T/A$. Thus, $K_T$ is also an index of fragility, which for obvious reasons we refer to as the {\it thermodynamic} fragility. Further, in making a comparison between kinetic and thermodynamic fragilities, we note that through the Adam-Gibbs relation, the relevant combination that is supposed to determine the kinetic fragility is $K_T/A$, which we have designated the {\it Adam-Gibbs fragility}, $K_{AG}$. In order to have a thermodynamic explanation of the fragility of a material, then, one would like to know $K_T$ and $A$, and be able to verify that the Adam-Gibbs fragility quantitatively matches the kinetic fragility $K_{VFT}$. 

The dependence of fragility of a glass former on the nature of inter-particle interactions is among the key aspects of fragility one would like to rationalise. Among studies addressing this aspect, a feature of focus has been how the softness of interactions may affect the fragility \cite{Mattsson,angellnv,affouard,sengupta2011,kelton,lwang}. In particular, Mattsson {\it et al} \cite{Mattsson} found experimentally that, with density as a control parameter, suspensions made of softer colloids had less fragile, or stronger, behaviour. However, computational studies using modified Lennard-Jones interactions, varying temperature at fixed density, Bordat {\it et al} \cite{affouard} found that the kinetic  fragility {\it increases} with increasing softness of interaction. In order to rationalise these observations, Sengupta {\it et al} \cite{sengupta2011} analysed the kinetic as well as thermodynamic fragilities. They found that the thermodynamic fragility $K_T$ decreases with increasing softness of interactions.  
{However}, consistently with Bordat {\it et al} \cite{affouard}, the kinetic fragility increases with softness, albeit modestly.  In seeking consistency between these results, Sengupta {\it et al} \cite{sengupta2011} evaluated the {\it Adam-Gibbs} fragilities, with the activation free energy $A$ in the AG relation being estimated from the high temperature activation energies. Indeed, the Adam-Gibbs fragilities agree rather well with the kinetic fragilities.  
{This result highlights} the importance of factoring in the activation free energy $A$ in analysing fragility. Sengupta {\it et al} \cite{sengupta2011} considered next an approach to understanding the variation of  $A$ in the AG relation employing the notion of density-temperature scaling, that has been explored in the study of dynamics in many glass formers, in particular the so-called strongly correlating liquids \cite{gnan2009}. In short, the premise is that the dynamics and thermodynamics of such liquids depends on a scaled variable $X \equiv T/\rho^{\gamma}$ where $\rho$ is the density of the liquid.  
{Various} ways of estimating $\gamma$ from correlations in the potential energy {\it etc} have been analyzed \cite{gnan2009}. The attempt in \cite{sengupta2011} to thus rationalise the model dependence of kinetic fragility using density-temperature scaling is only modestly successful. On the other hand, the same approach is successful in rationalising the density dependence of fragilities in the same model \cite{sengupta2013}. Thus, while the use of density-temperature scaling is a promising way to analyse the dependence of fragility on softness of interactions, the analysis thus far remains incomplete. Further, all the theoretical analyses referred to use temperature as the control parameter with respect to which the changes in dynamics are studied, whereas the experimental study mentioned above \cite{Mattsson} used density as the control variable. It is thus interesting to study the dynamics of the series of model liquids mentioned above varying the density rather than temperature to obtain estimates of kinetic fragility that may then be analysed through thermodynamic characterisation {\it via} the configurational entropy. We perform such an analysis here by performing computer simulations of the model liquids varying the density at fixed temperature. We  
 analyse the fragility employing the inverse density in place of temperature as the control parameter as done in experiments to obtain the expected trend in fragility with softness. We then perform analysis to examine whether density-temperature scaling holds in these systems, and if so with what parameters $\gamma$. We  
 proceed to define the scaled parameter $X$, and perform further analysis with this scaled variable. 

If one assumes that the  dynamics and thermodynamics of the system is controlled by the scaled variable $X$, it is straight forward to write the corresponding relations that we should employ, by assuming that $T_{VFT}$, $T_{K}$, $K_{T}$ and $A$ have density dependences, such as $T_{VFT} = X_{VFT} \rho^{\gamma}$. From these we write the $X$ dependent equations

\begin{equation}
B(X) = B_{0} \exp\left[\frac{1}{K^{(X(B))}_{VFT}\left (\frac{X}{X_{VFT}}-1 \right )} \right] 
\label{Xfragility}
\end{equation}

\begin{equation}
B(X) = B_{0} \exp\left[\frac{A^{X\, (B)}}{XS_{c}}\right]
\label{eqn:AG}
\end{equation}

\begin{equation} \label{eq:kagPEL}
X S_c(X) = K^{X}_{T}~~ \left ({X \over X_K} -1 \right), 
\end{equation}\label{XScvsX}

where we have added a superscript $X$ to $K_{VFT}$,  $K_T$ and $A$ to indicate that these values are different from the ones that appear in the $T$ dependent equations above. Thus, we first analyse the dynamical quantities using the $X$ dependent VFT relation to obtain the kinetic fragilities and the glass transition value $X_g$. We next consider the $X$ dependence of $X S_c$ to obtain the thermodynamic fragility. In the simplest scenario, the thermodynamic fragility must equal the kinetic fragilities. However, the dependence of $A^X$ on the model parameters, or indeed additional dependence of density that may not be captured by density-temperature scaling, cannot be excluded, and thus, we also consider estimating the density dependence directly from the relationship between dynamical quantities and the configurational entropy. 

The remaining paper is organized as follows: In the next section, we describe the models and methods employed for our analysis. In Sec. 3, we present our results. In Sec. 4, we discuss the conclusions from our analysis and summarise our findings.

\section{Models and Methods}
We have performed molecular dynamics simulations for the binary mixture  ($A_{80}B_{20}$), interacting {via} Lennard-Jones potential in three dimensions \cite{KAref}. The interaction potential is
\begin{eqnarray}
V_{\alpha\beta} &=& \frac{\epsilon_{\alpha\beta}}{q-p}\left[ p\left( \frac{r_{\alpha\beta}^{min}}{r}\right)^q - q\left(\frac{r_{\alpha\beta}^{min}}{r}\right)^p \right] \nonumber  \\ 
 & & +  c_{1~ \alpha\beta}r^2 + c_{2~ \alpha\beta}, \quad  r_{\alpha\beta}<r_{c,\alpha\beta}=2.5\sigma_{\alpha\beta} \nonumber \\
 &=& 0,\quad \text{otherwise,} 
\end{eqnarray}
where, $\alpha\, \beta \in [A,B]$.  The truncation coefficients $c_{1~\alpha\beta}$ and $c_{2~\alpha\beta}$ make the potential and force  go to zero smoothly at cutoff ($r_{c,\,\alpha\beta}$) and $r_{\alpha\beta}^{min} (= 2^{1/6} \sigma_{\alpha\beta})$ is the value at which pair potential is minimum. Length, temperature and time are measured in units of $\sigma_{AA}$, $\epsilon_{AA}/k_B$ and $(m_{AA} \sigma_{AA}^2/\epsilon_{AA})^{1/2}$, respectively. In reduced units, $\epsilon_{BB} = 0.50$ and $\epsilon_{AB} =1.5$ $\sigma_{BB} = 0.88 $ and $\sigma_{AB} =0.80$. The softness of the interaction can be tuned by the choice of the different values for [q, p]. We study the models  [12,11], [12,6] and [8,5] for temperatures $T= 1.0, 0.8$ and $0.6$,  for a range of densities in each case. We perform constant temperature molecular dynamics simulations for $N=1000$ particles with time step $dt= 0.001$.
 To keep  temperature constant we use the Brown and Clarke algorithm \cite{BC}. 
The simulations were performed for densities 
$\rho \in (1.05,1.23)$, $(1.10,1.23)$ and $(1.19, 1.845)$ for the [12,11], [12,6] and [8,5] models respectively. The simulation run lengths were  longer than $100\tau$ (described below). The lower limit of the density for  
{each model} is chosen to be above the  density at which the pressure of the inherent structures display minima\cite{sengupta2011}.  

We characterize the dynamics by evaluating the diffusion coefficient and relaxation times of the $A$ type particles. The diffusion coefficient $(D_{A})$ is estimated from the mean squared displacement (MSD). The relaxation times are obtained from the decay of  self intermediate scattering function ($F_s(k,t)$) using the definition $F_s(k\sim7.25,\tau) = 1/e$.The self  intermediate scattering function is defined as 
\begin{equation}
F^{AA}_s(\vec{k},t) = \frac{1}{N_A}\left<\sum_{i=1}^{N_A} exp(- i\vec{k}.(\vec{r}_{i}(t)-\vec{r}_{i}(0)))\right>
\label{scattfunc}
\end{equation}
where $\vec{r}_{i}(t)$ is the position of particle $i$ at time $t$, and ``$<>$" represents the average over time origins.  

The configurational entropy is calculated by subtracting  the  vibrational component of the entropy from the total entropy~\cite{sastry2000}:
\begin{equation}
S_c(T) = S_{total}(T) -S_{vib}(T)
\label{confent}
\end{equation}
The total entropy of the liquid structure estimated from the thermodynamic integration, using ideal gas limit as the reference state. The vibrational part of the entropy is estimated by approximating each basin as a harmonic well ~\cite{sastry2000, sastry2000a}.  

\section{Results}
The results from the simulations regarding density-temperature scaling, the model dependence of the kinetic and thermodynamic  fragilities  (the latter being evaluated from the variation of configurational entropy) and the Adam-Gibbs fragilities from analysis of the Adam-Gibbs relation, are described below. 

\subsection{Density temperature (DT) scaling}
The exponent $\gamma(\rho,T)$ can be computed from the fluctuation in the potential energy ($U$) and virial ($W$)\cite{Pedersen2010},
\begin{equation}
\gamma = \frac{\left<\Delta W \Delta U \right >}{\left<(\Delta U)^2\right>},
\label{DTgamma}
\end{equation}
where $\Delta U = U - <U>$ and  $\Delta W = W - <W>$ are the spontaneous fluctuations in the potential energy and virial about their means values. The angular brackets ``$\left< \right >$" represents the ensemble average of the fluctuations.
In Fig. \ref{fig:gamma} we show the scaling exponent is obtained from the ratio of the fluctuations in energy and the correlated fluctuations of the energy and the virial.  We find that the exponent has values: $6.82 \pm 0.1$, $5.28 \pm 0.1$ and $3.61 \pm 0.1$ for model [12,11], [12,6] and [8,5] respectively, for temperatures and densities as shown in Fig. \ref{fig:gamma} (a). The error bars
{are} obtained from computing averages for different blocks of the trajectory, and then obtaining the mean and variance of these block averages. Fig. \ref{fig:gamma} (b) shows the variation of the $\gamma$ values with density at $T = 1$. According to thermodynamic theory of DT scaling \cite{gnan2009,gnan2011}, both relaxation time and configurational entropy show DT scaling: $\tau = f(\rho^\gamma/T)$ and $S_c = h(\rho^\gamma/T)$.  We verify  (Fig. \ref{scaling}) DT scaling for the studied models for the reduced relaxation time ($\tau^*  = \rho^{1/3} (k_BT/m)^{1/2}\tau$) and observe that the scaling is good in the range of densities and temperature studied. The scaled variable $\rho^\gamma/T$  captures the density and temperature variation of the relaxation times in the liquids. 
A given value of the relaxation time can be obtained for different combinations of temperature and density, if they are changed in a correlated way so as to keep the scaled variable $\rho^\gamma/T$ constant.
 For analysis to be discussed later, we also compute the exponent $\gamma$ from the Adam-Gibbs (AG) relation, {i.e.}  $TS_c ln(B/B_0)= A(\rho)$, where $ A(\rho) = A_o^{(q,p)} \rho^{-\gamma}$. $B$ is either the diffusion coefficient or the relaxation time \cite{sengupta2013} (Fig. \ref{DT:AGgamma1})  
. As shown in Fig. \ref{DT:AGgamma1},   and tabulated in Table \ref{tableX1}, the $\gamma$ values so obtained have the same trend as those mentioned earlier in terms of their model dependence, but are on average higher in value. In the present study we characterise the fragilities in terms of the scaled variable  $\rho^\gamma/T$, in addition to using the inverse density as the control parameter, as we discuss next. 

\begin{table}[h]
\centering
 \begin{tabular}{||c c c c||} 
 \hline
 Model  &   $D_o$ & $A_o^{(q,p)} $ & $\gamma$      \\
 [0.5ex] 
 \hline \hline
 12,11 	& 0.0412	&-0.49499	& 8.2049	\\
 12,6 	& 0.0622	&-0.494784	& 5.30077   \\
 8,5 	& 0.0596	&-0.268537	& 4.36905   \\ [1ex] 
 \hline
 \end{tabular}
 \caption{Fit parameters for  $TS_c ln(D/D_0)=  A_o^{(q,p)} \rho^{-\gamma}$. $D$ values used are not scaled.}
 \label{tableX1}
\end{table}

\begin{figure}[h!]
\centering
\includegraphics[scale=0.29]{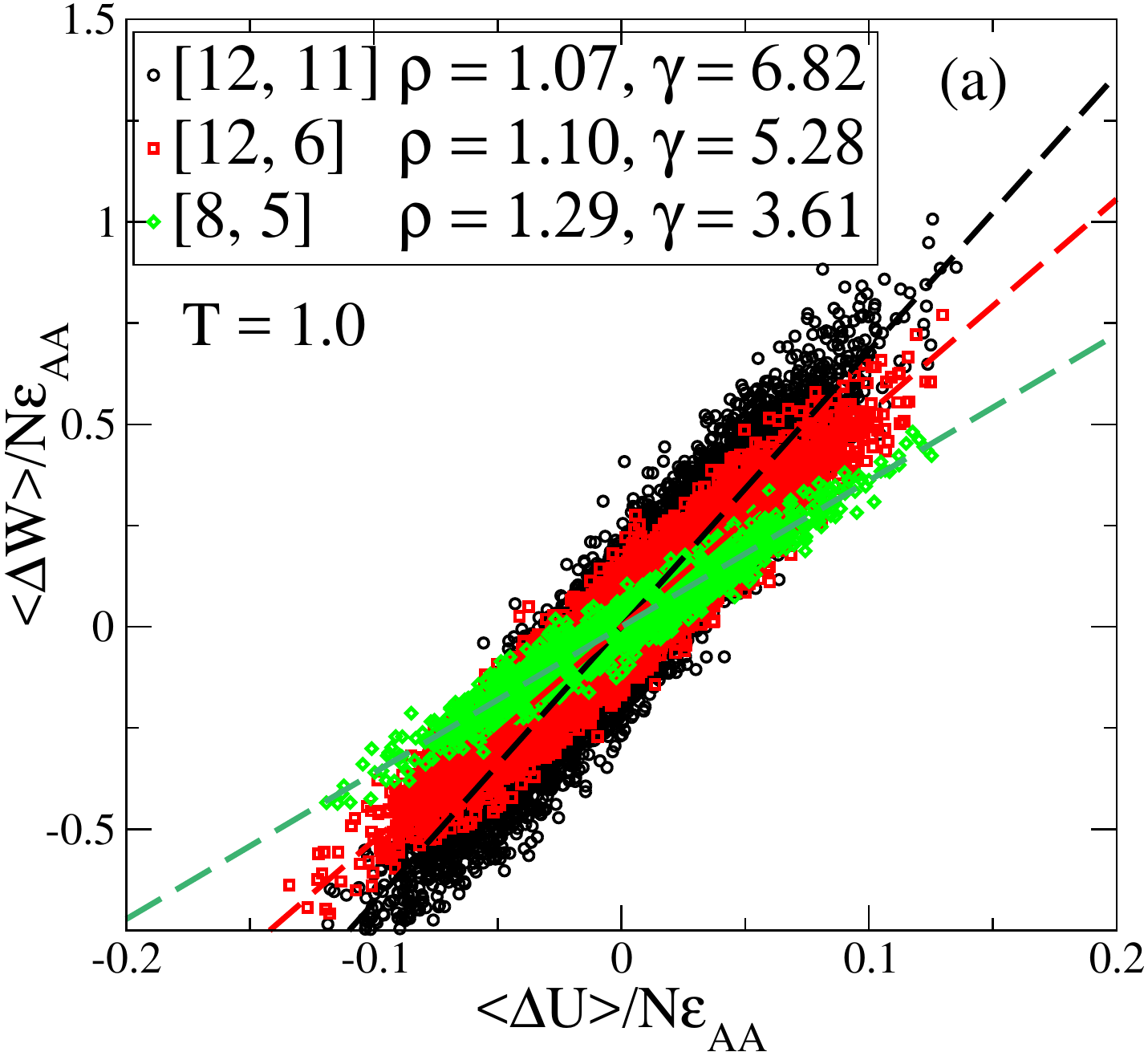}
\includegraphics[scale=0.29]{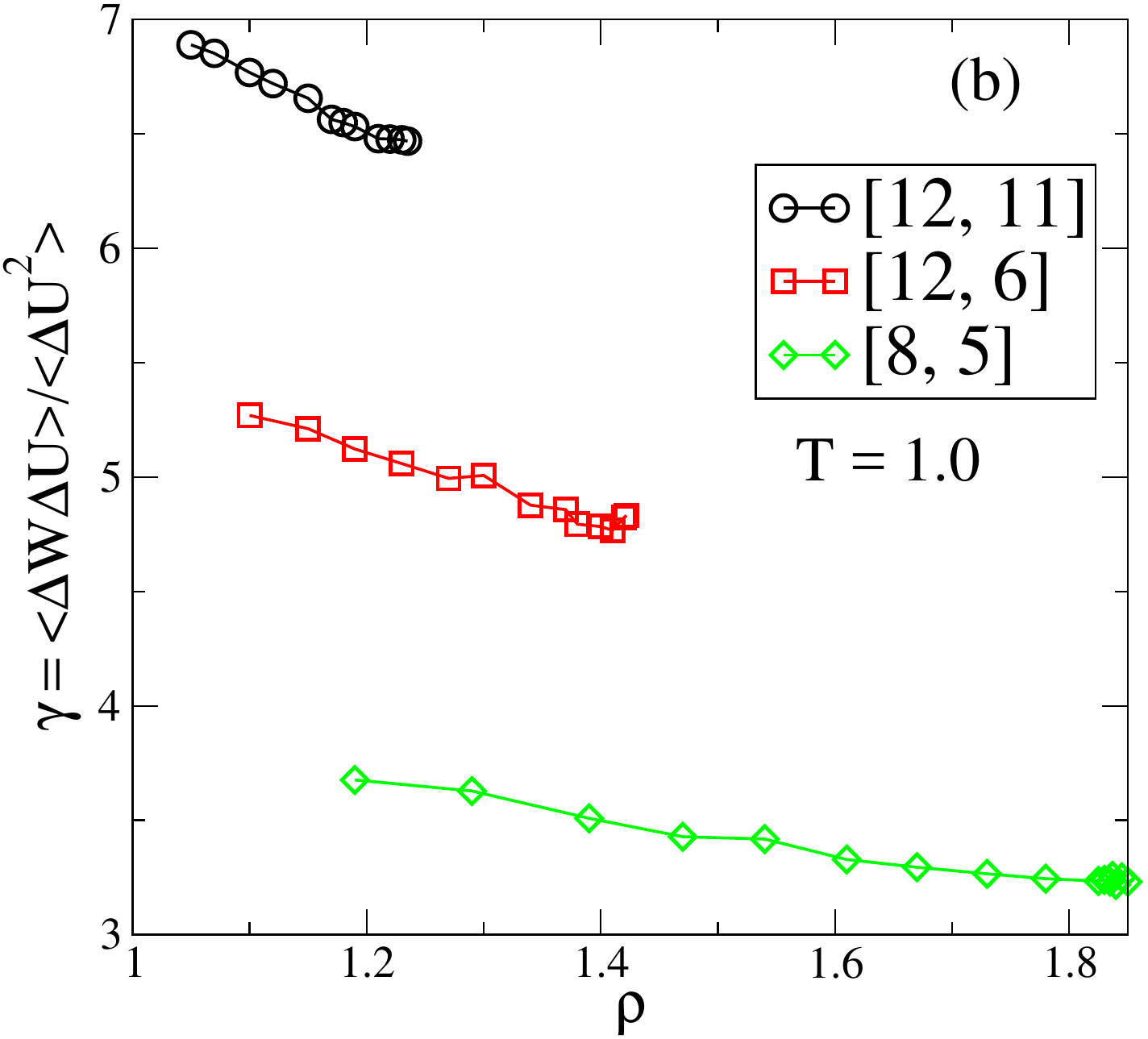}
\caption{Panel (a) illustrates the estimation of the DT scaling exponent ($\gamma$) for the correlation from the instantaneous energy ($\Delta U = U - \left <U\right>$) and virial ($\Delta W = W - \left <W\right>$). (b) The density dependence of the scaling exponent $\gamma$ for the studied models.}
\label{fig:gamma}
\end{figure}

\begin{figure}[h!]
\centering
\includegraphics[scale=0.35]{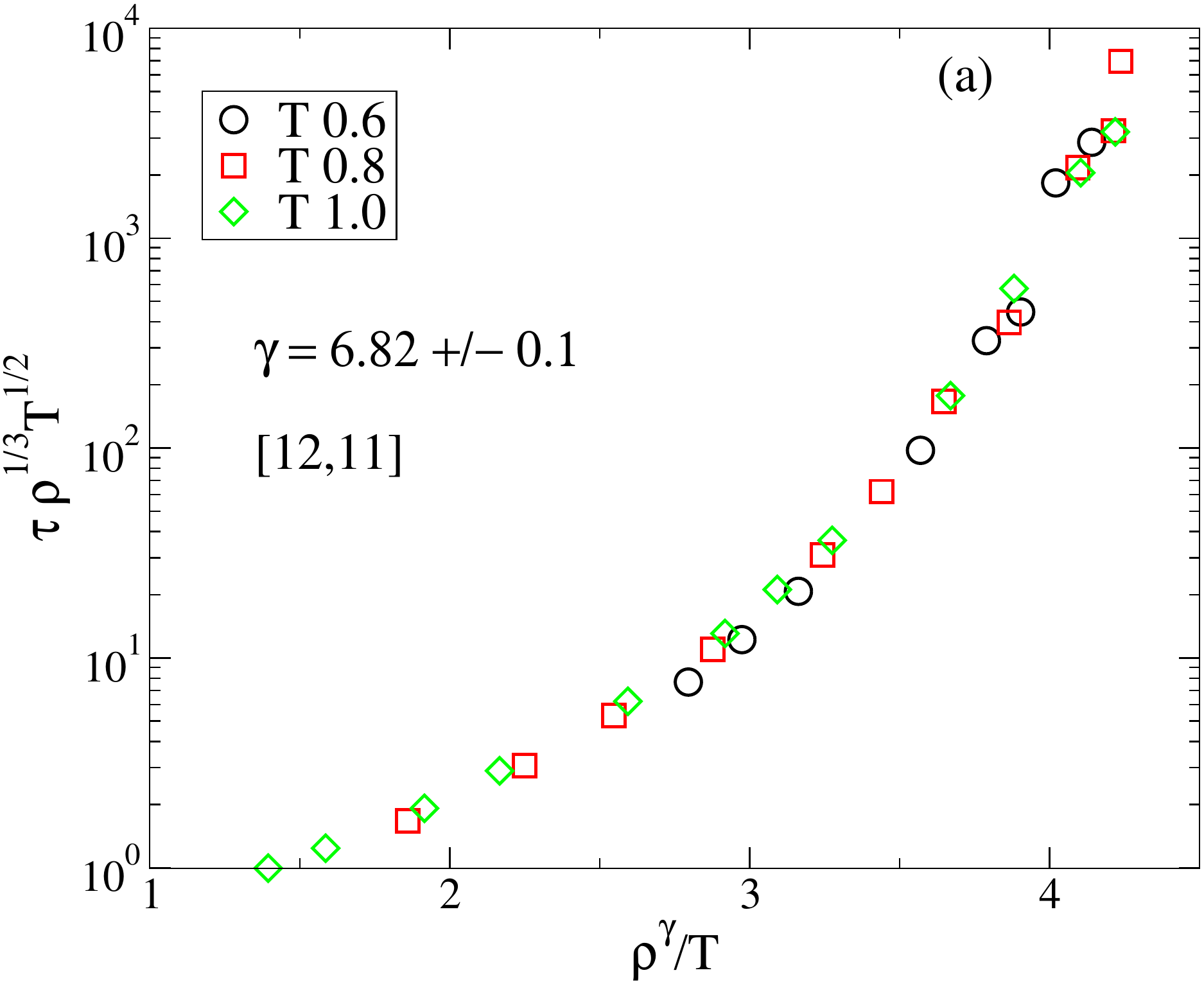}
\includegraphics[scale=0.35]{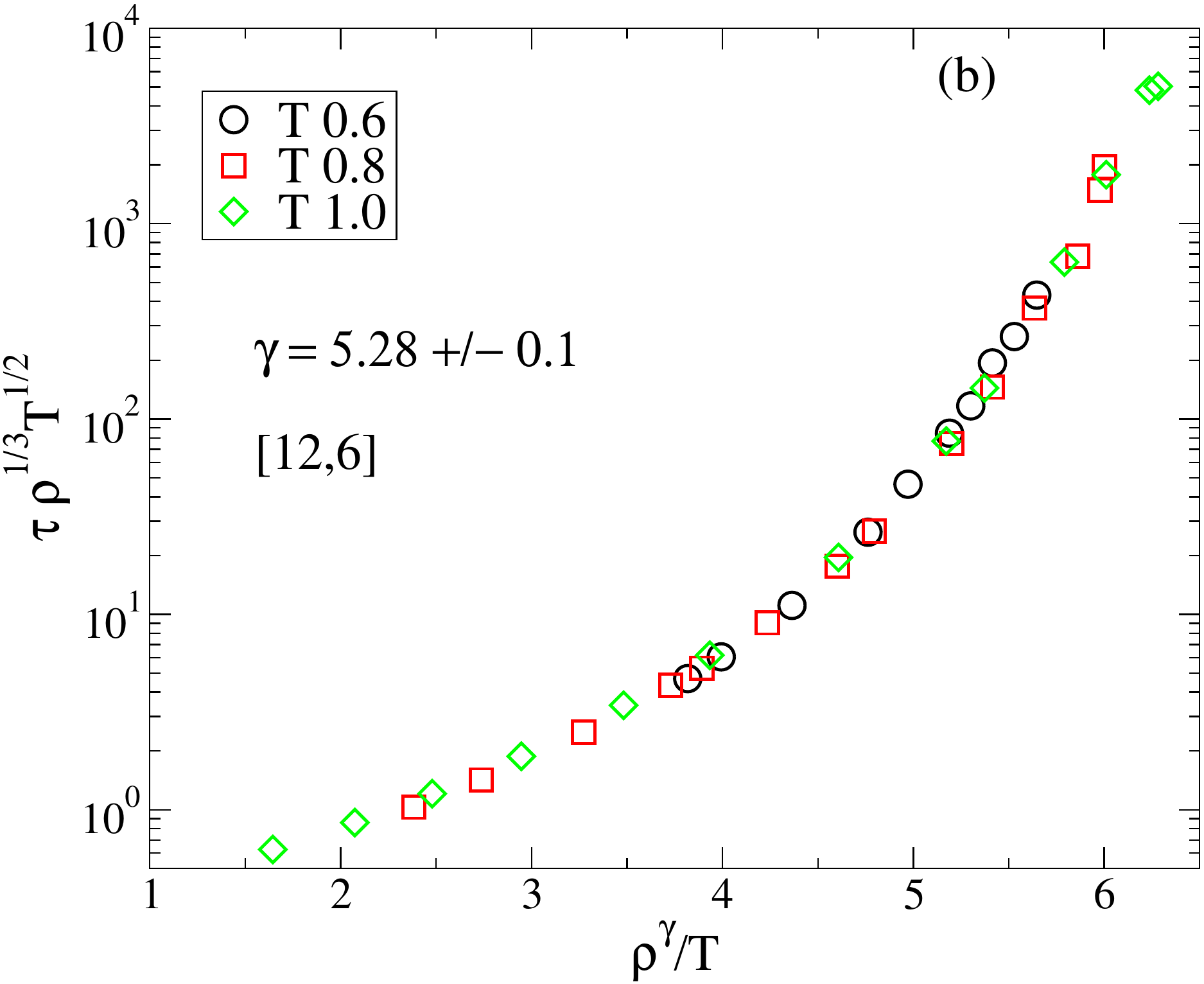}
\includegraphics[scale=0.35]{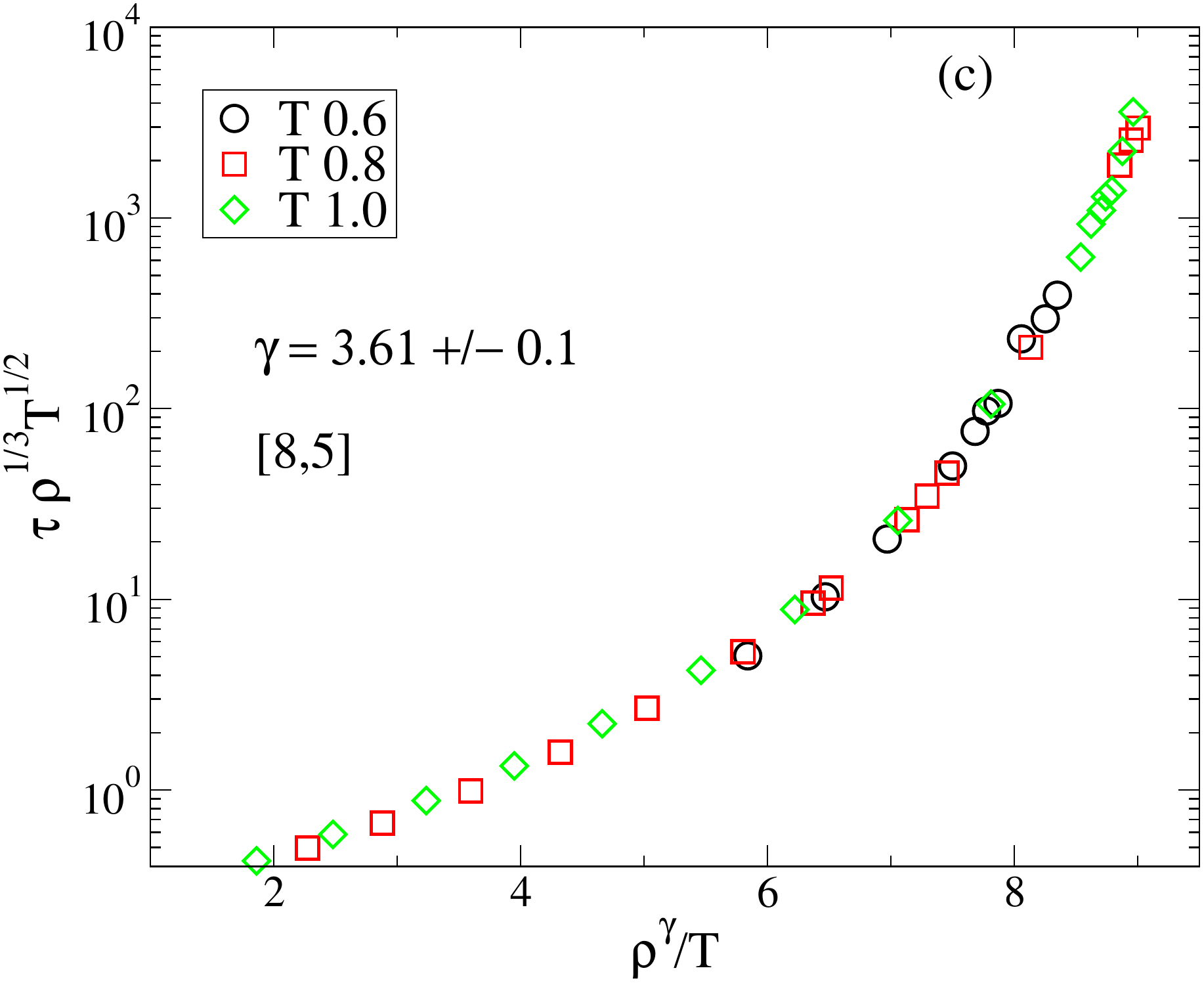}
\caption{The density temperature scaling for various models for the relaxation time $(\tau)$ of particle type $A$. The DT scaling exponents are estimated for the Eq. \ref{DTgamma} are $6.82$, $5.28$ and $3.61$ for the models [12,11], [12,6] and [8,5] respectively.}
\label{scaling}
\end{figure}

\begin{figure}[h!]
\centering
\includegraphics[scale=0.32]{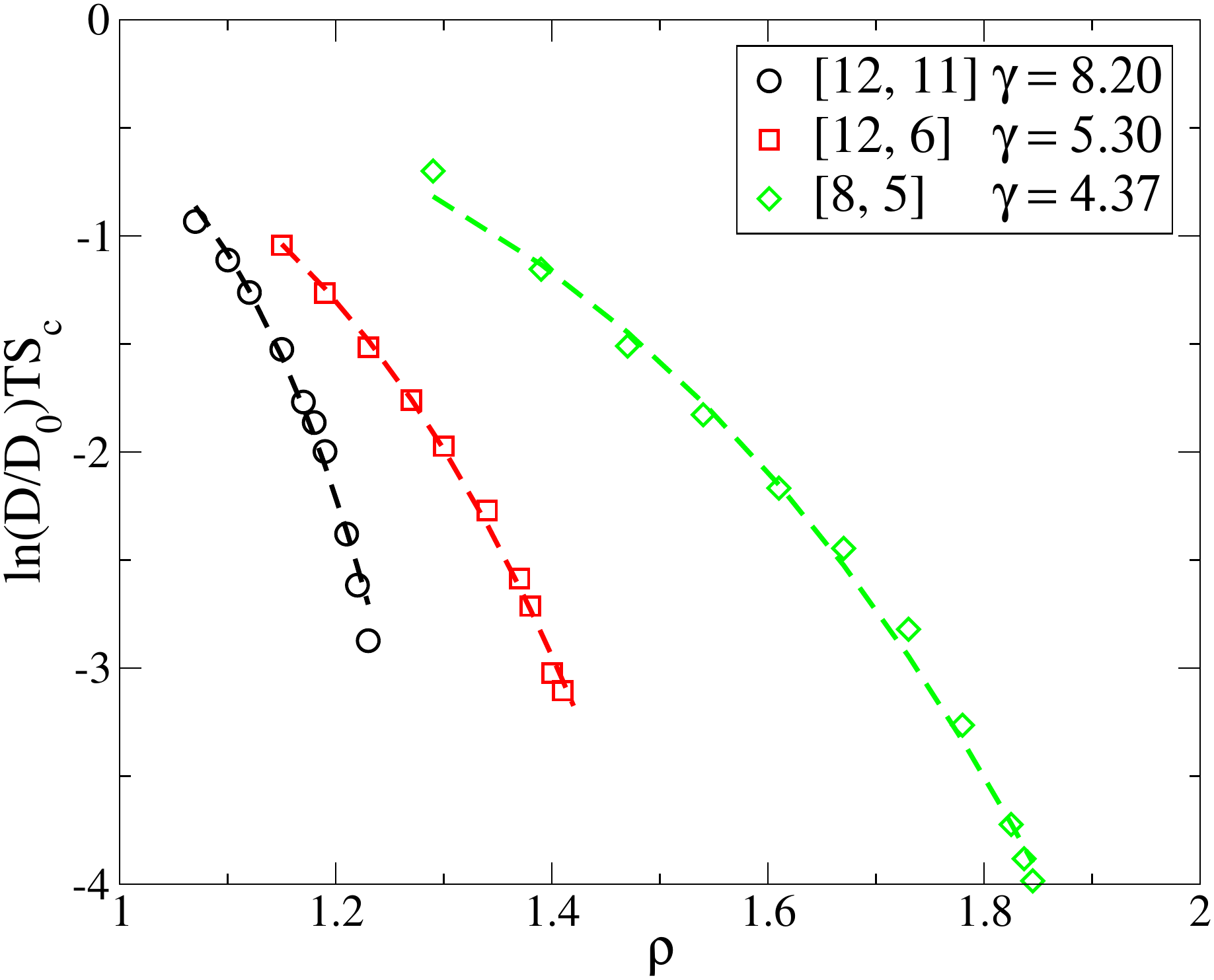}
\caption{The DT scaling exponents $\gamma$  estimated {\it empirically} using {i.e.} $TS_c\,ln(D/D_0) = A_0^{(p,q)}\rho^{-\gamma}$.}
\label{DT:AGgamma1}
\end{figure}

\begin{figure}[h!]
\centering
\includegraphics[scale=0.28]{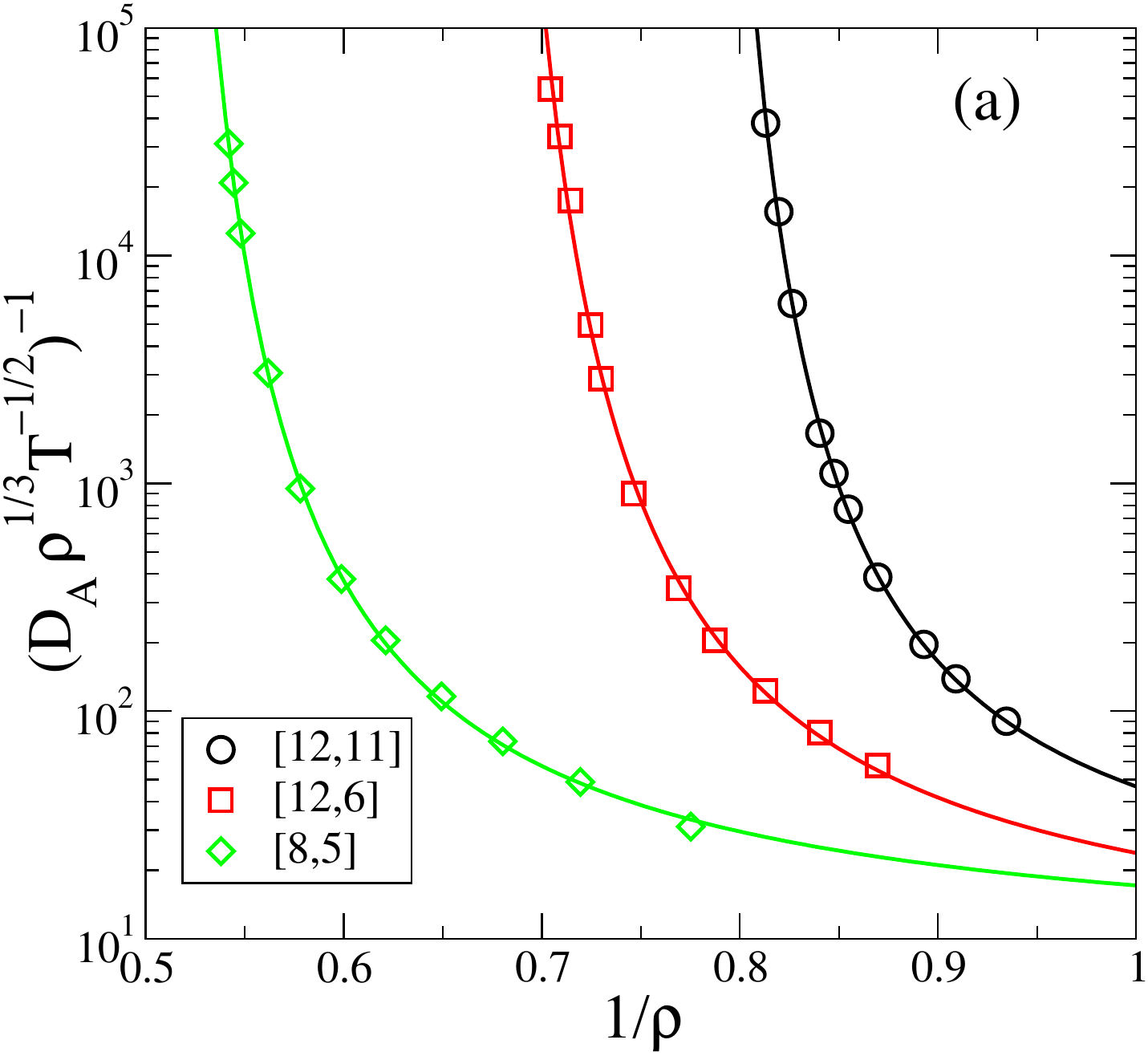}
\includegraphics[scale=0.28]{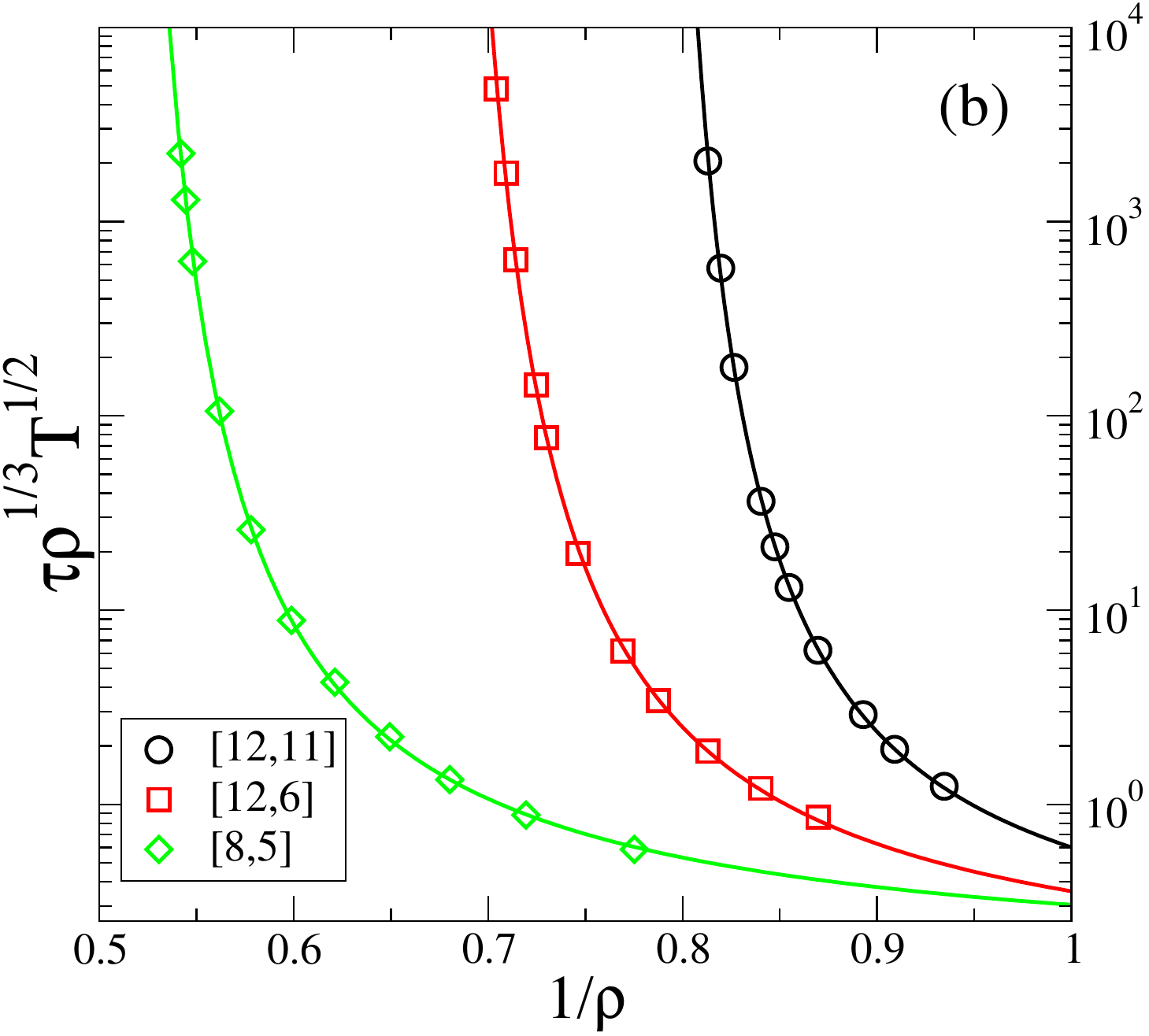}
\includegraphics[scale=0.28]{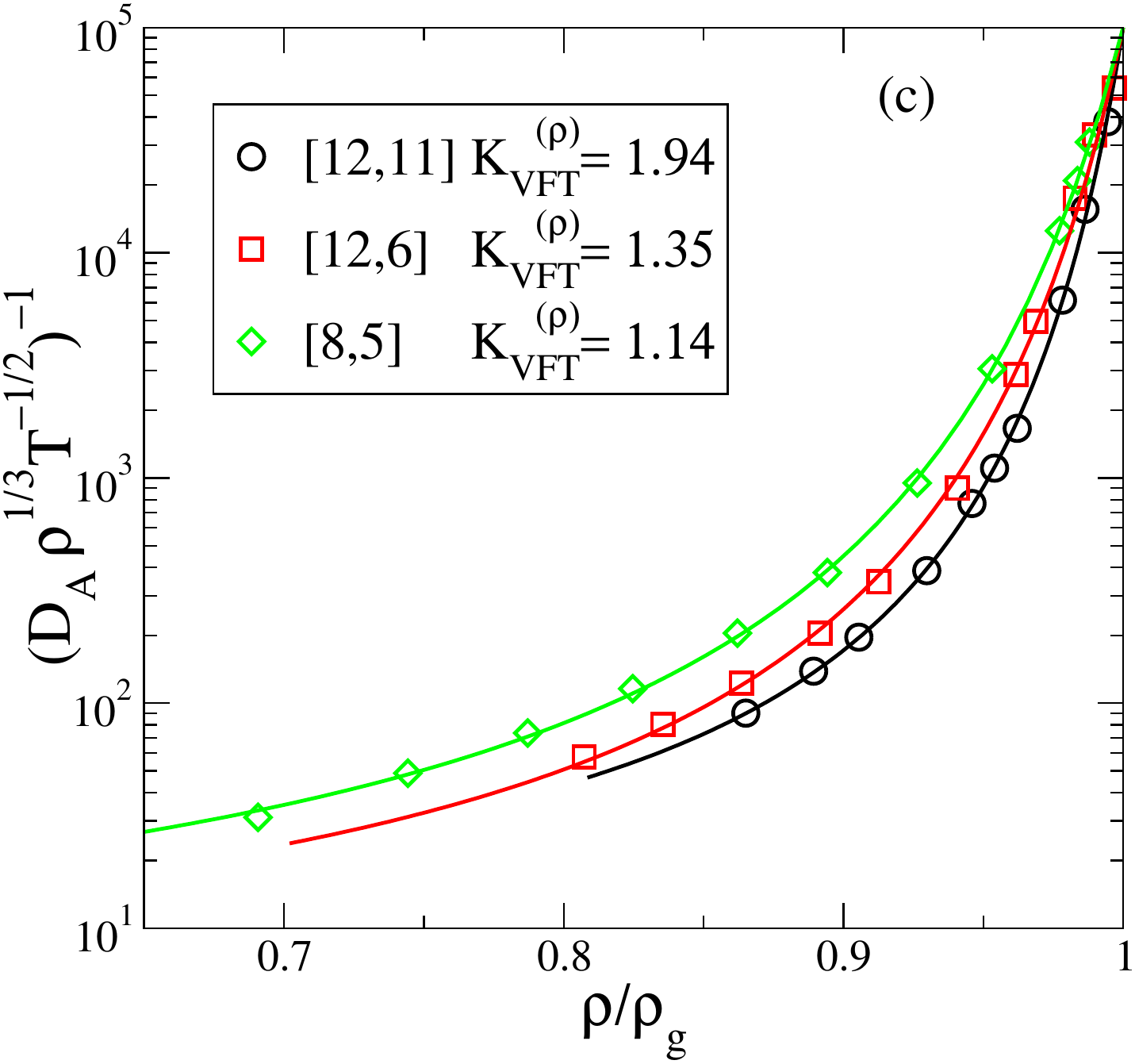}
\includegraphics[scale=0.28]{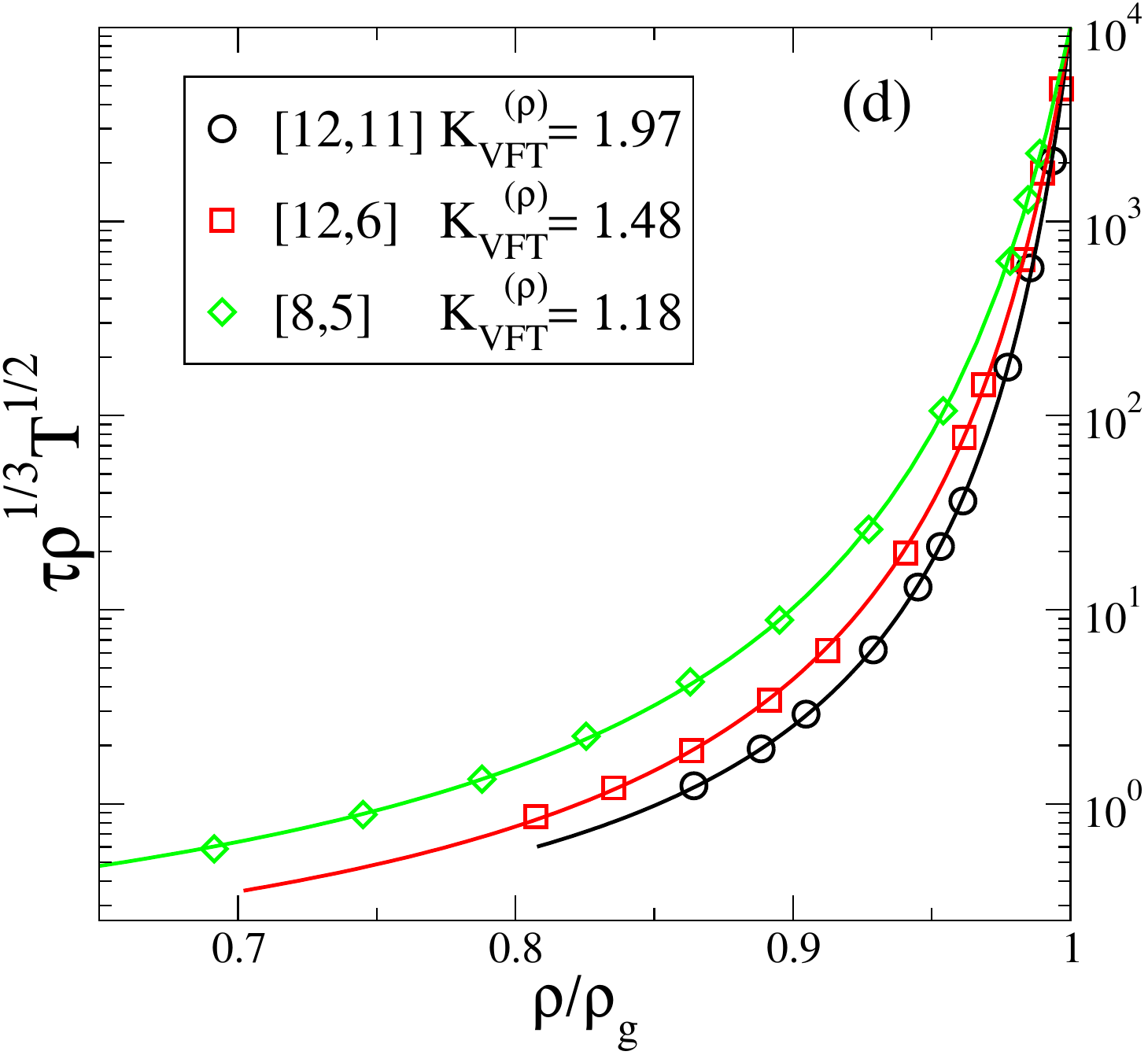}
\caption{(a) The inverse diffusion coefficient and (b) relaxation time as a function of the inverse density for the studied models. The lines through the symbols are VFT fit to the simulation data. The Angell plot for (c) the diffusion coefficient and (d) relaxation times {\it vs.} scaled density $\rho/\rho_g$. The density $\rho_{g}$ corresponds to the value at which inverse diffusion coefficient and relaxation time become, $10^5$ and $10^4$ respectively. The kinetic fragilities $K_{VFT}^{(\rho)}$, estimated from the VFT fits, indicate that softer interactions correspond to stronger glass formers, consistently with the trend observed by Mattsson {et al.}\cite{Mattsson}. }
\label{fig:invrgo_VFT_ANGELL}
\end{figure}

\begin{figure}[]
\centering
\includegraphics[scale=0.285]{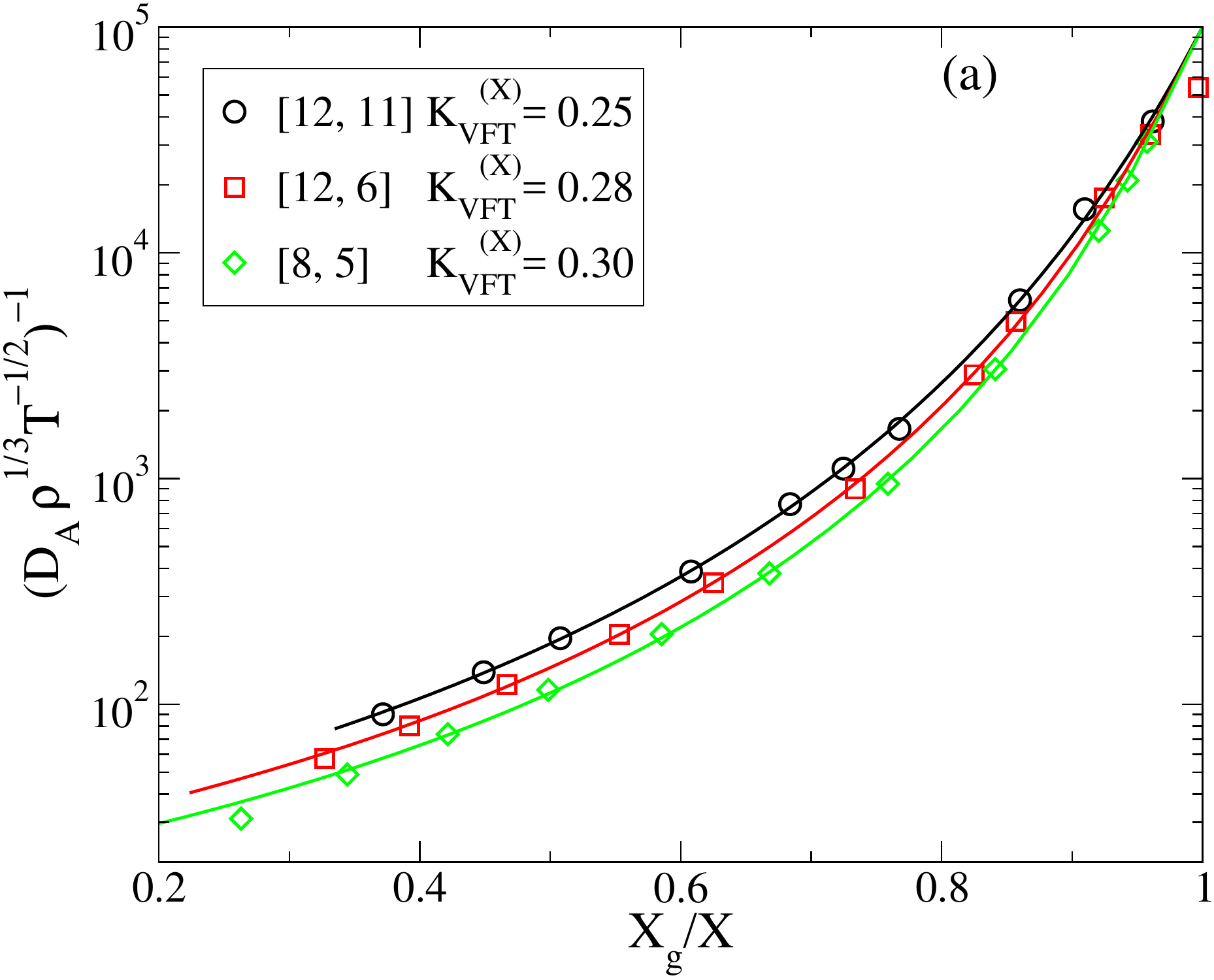}
\includegraphics[scale=0.285]{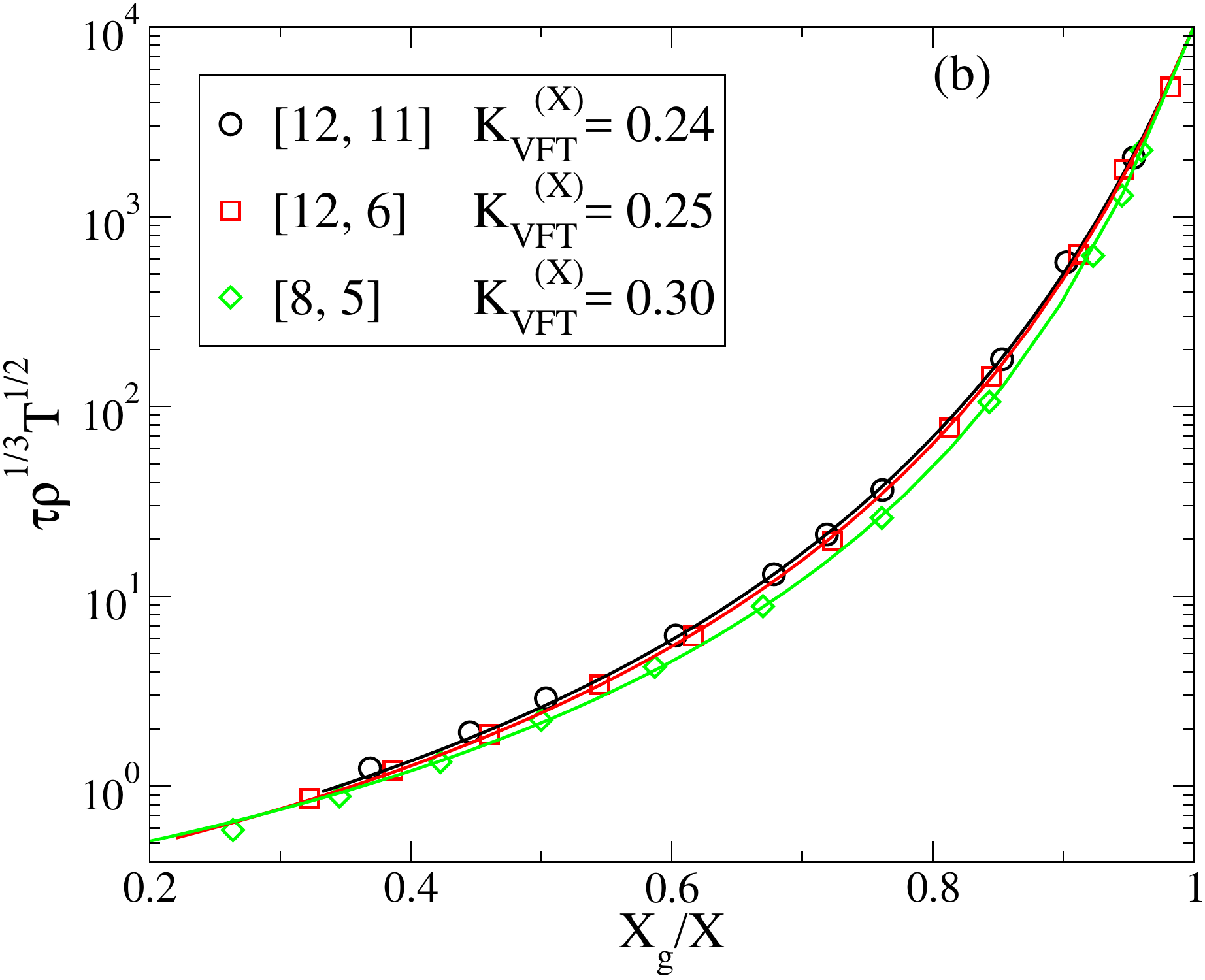}
\caption{The Angell plot for (a) the diffusion coefficient and (b) the relaxation time with  the variable $T/\rho^\gamma (=X)$ for the studied models. The variable $X_g$ corresponds to the value at which inverse diffusion coefficient is $10^5$ and relaxation time becomes $10^4$. The kinetic fragility $K_{VFT}^{(X)}$ although showing weak dependence on the softness of interactions, increases with increasing softness, and is inconsistent with the trend observed by Mattsson {et al.}\cite{Mattsson}.}
\label{Angell_X_gamma}
\end{figure}

\subsection{Kinetic fragility}
We analyse the density dependence of the relaxation time and diffusion coefficient for $T=1.0$ to estimate kinetic fragilities.  The kinetic fragility is estimated from the variation of relaxation times and diffusion coefficients with inverse density and the scaled variable $(X\equiv \rho^\gamma/T)$ as control parameters.

\begin{table}[h]
\centering
 \begin{tabular}{||c c c c c||} 
 \hline
  & Model & $1/D_o$ or $\tau_o$ &  $K_{VFT}^{(\rho)}$ & $1/\rho_{VFT}$ \\ [0.5ex] 
 \hline \hline	
$\tau$        &12,11	& 0.10621	&1.96861	&0.773432 \\
			   &12,6	& 0.09454	&1.48144	&0.663272\\	
			   &8,5		& 0.13084	&1.17694	&0.498354	\\
$1/D_A$		   &12,11	& 8.62444	&1.94229	&0.766331\\
			   &12,6	& 5.93977	&1.35195	&0.652424\\
	           &8,5		& 7.36353	&1.13827	&0.490223\\[1ex] 
 \hline	          
\end{tabular}
\caption{Fit parameters for VFT fits using $1/\rho$ as the control parameter. Scaled $D_A$ and $\tau$ values are used.}
\label{tableX2}
\end{table}

In Fig. \ref{fig:invrgo_VFT_ANGELL} (a-b), we show the inverse of the scaled diffusion coefficient $(D_{A})$ and relaxation times $(\tau)$, plotted against the inverse density ($1/\rho$). The lines show the VFT fits for these dynamical quantities, using $1/ \rho$ instead of $T$ in Eq. \ref{eqn:kinfr2}. Fit parameters for these VFT fits are tabulated in Table \ref{tableX2}.
The VFT fits for the $1/\rho$ provides estimates of the kinetic fragility $K_{VFT}^{(\rho)}$ and relaxation time divergence density $\rho_{VFT}$, which are listed in the Table \ref{table1}, which we discuss later.

In the Fig.  \ref{fig:invrgo_VFT_ANGELL} (c-d), the we show the Angell plot, 
{i.e.} dynamical quantities plotted against scaled density $\rho/\rho_g$, where $\rho_g$ is the value of the density (extrapolated using the VFT form) at which inverse diffusion coefficient and relaxation time reach values of $10^5$ and $10^4$ respectively.   

We find that the kinetic fragility calculated as described decreases as the interparticle interactions become softer, which is consistent with the finding of Mattsson {et al.} \cite{Mattsson} but shows a opposite trend as reported by Sengupta {et al.} \cite{sengupta2011}, using the temperature variation of dynamical quantities. 

Next we measure the kinetic fragility  using the variation of dynamical quantities considering their variation with the scaled variable $X$, using Eq. \ref{Xfragility}.  We show in Fig. \ref{Angell_X_gamma} the Angell plots {\it vs.} $X$, and the corresponding VFT fits, whose fit parameters are listed in Table \ref{tableX3}.
The kinetic fragility $K_{VFT}^{(X)}$ using $X$ as the control parameter increases as the softness of interactions increases, which is inconsistent with results reported by Mattsson {et al.} \cite{Mattsson}.  

\begin{table}[h]
\centering
 \begin{tabular}{||c c c c c||} 
 \hline
  & Model & $1/D_o$ or $\tau_o$ &  $K_{VFT}^{(X)}$ & $X_{VFT}$ \\ [0.5ex] 
 \hline \hline	
$\tau$        &12,11	& 0.255324 &	0.24156 &	0.167089 \\
			   &12,6	& 0.249459 &	0.250601&	0.112048\\	
			   &8,5		& 0.283228 & 0.301737	& 0.0799073	\\
$1/D_A$		   &12,11	& 24.26900 & 0.251465	& 0.158602\\
			   &12,6	& 20.72685 & 0.277721 & 0.109851\\
	           &8,5		& 16.77903 & 0.296110 & 0.0755509\\[1ex] 
 \hline	          
\end{tabular}
\caption{Fit parameters for VFT fits using $X$ as the control parameter. Scaled $D_A$ and $\tau$ values are used. }
\label{tableX3}
\end{table}

\subsection{Thermodynamic fragility}

In Fig. \ref{fig:therofrag}(a) we show the configurational entropy plotted as  $XS_c$ {\it vs.} $X$, which shows that the data for the most part, especially at higher densities, obey the linear for expected (see Eq. \ref{XScvsX}). The linear behaviour permits the estimation of $K_{T}^{(X)}$ and the $X_K$ (analogous to the Kauzmann temperature) at which the configurational entropy vanishes by extrapolation. The thermodynamic fragility defined as Eq. \ref{XScvsX}, listed in Table \ref{tableX4}, is found to decrease as the interparticle interaction becomes softer, as displayed in Fig. \ref{fig:therofrag}(b).

\begin{table}[h]
\centering
 \begin{tabular}{||c c c c||} 
 \hline
   & Model & $K_T^{(X)}$ &  $X_{K}$  \\ [0.5ex] 
 \hline \hline	
&12,11	& 0.1099	   &	0.132784  \\
&12,6	& 0.09572	   &    0.0959724	\\	
&8,5		&  0.08322 &	0.0652737\\ [1ex] 
 \hline	          
\end{tabular}
\caption{Thermodynamic fragilities and Kauzmann points from fits to $XS_c$  vs. X}
\label{tableX4}
\end{table}

\begin{figure}[h!]
\centering
\includegraphics[scale=0.310]{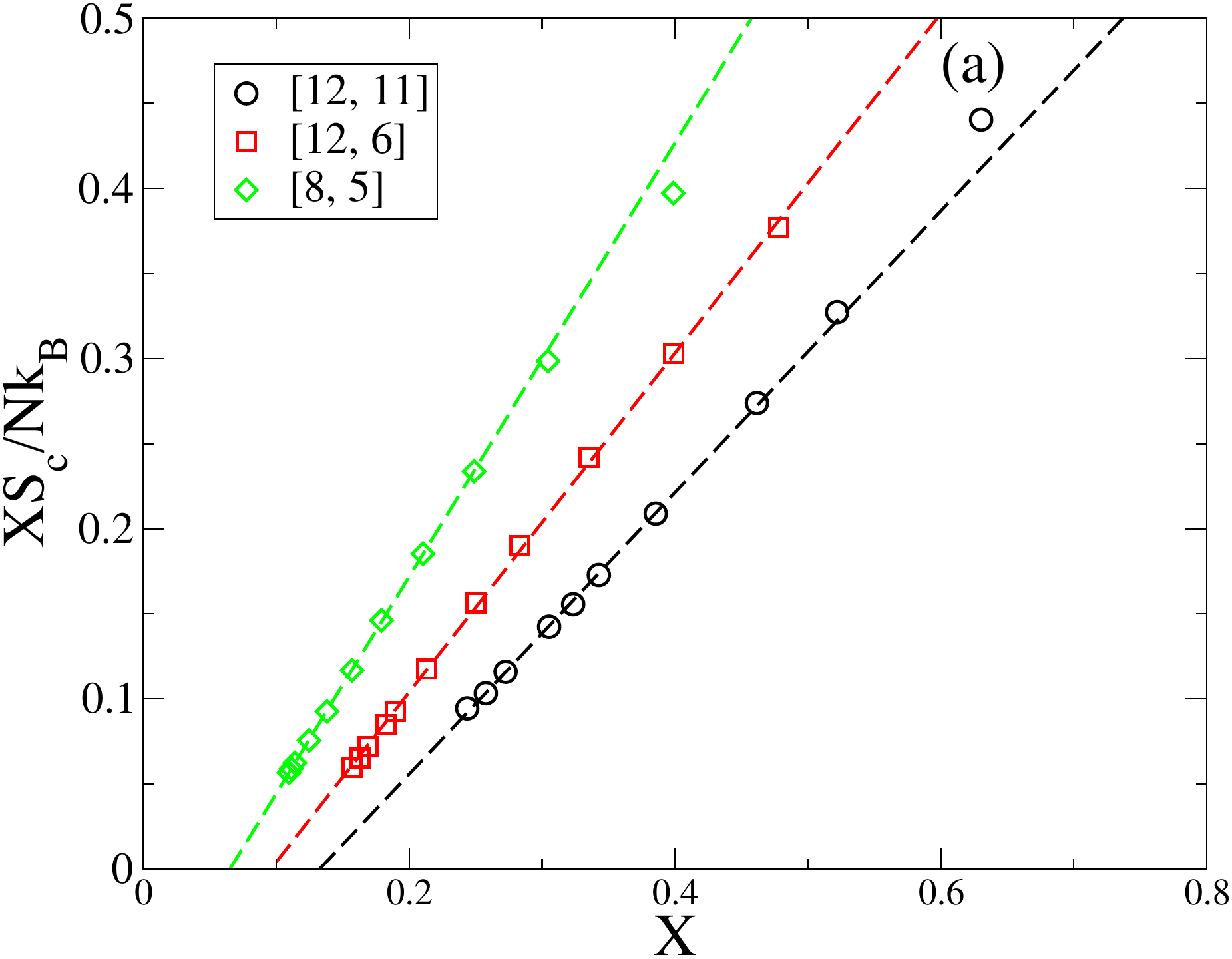}
\includegraphics[scale=0.310]{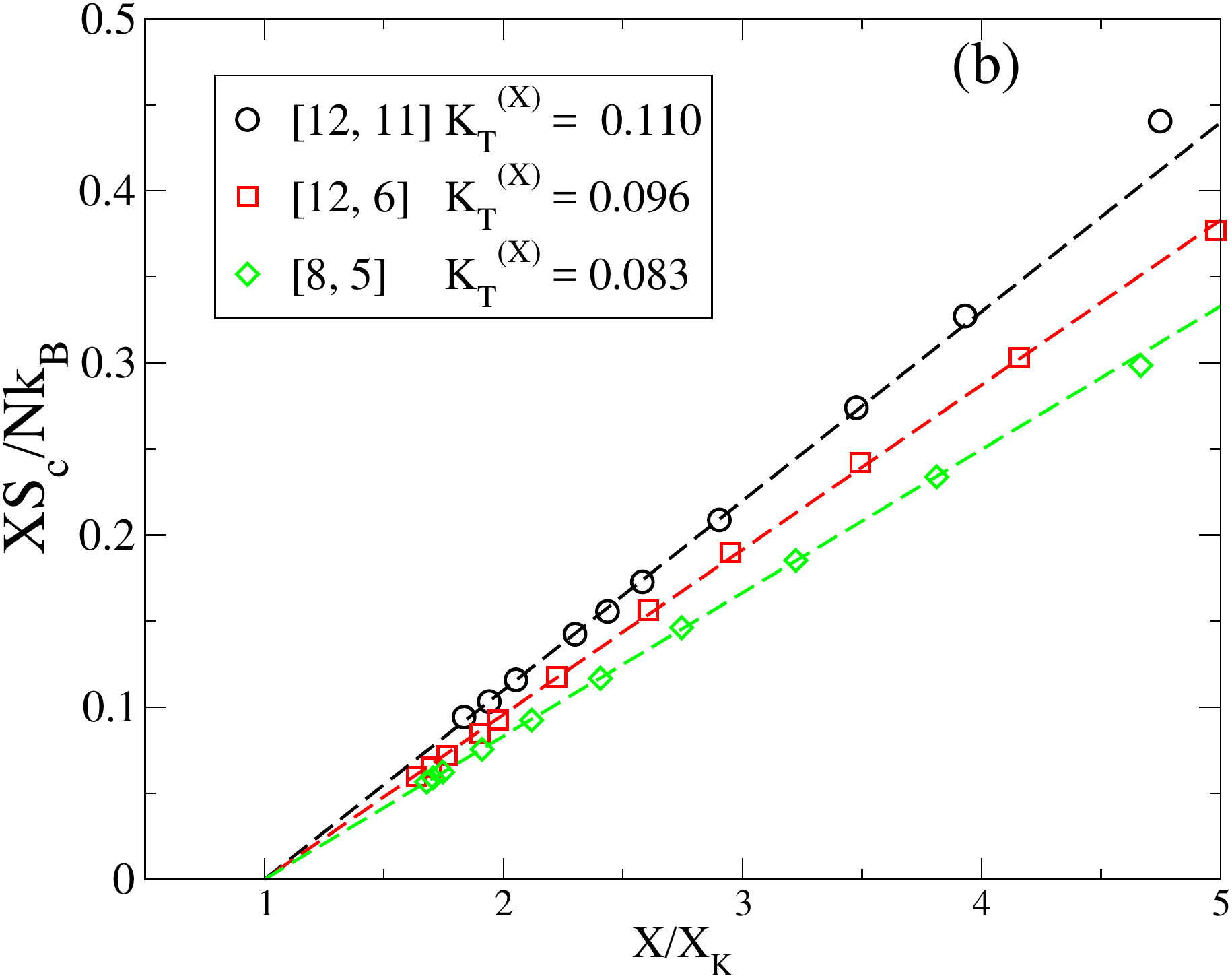}
\caption{{(a)} The configurational entropy $S_c$ is estimated by subtracting  out the vibrational component of the entropy from the total entropy.  $XS_c$ has been plotted against parameter $X$, and lines are linear fits to the data as defined in Eq. \ref{eq:kagPEL}. The  $X_K$ (analogous to Kauzmann temperature), at which $S_c=0$, is $X_K =~0.13,~0.10,~0.07$, for model [12,11], [12,6] and [8,5], respectively.  
{(b)}  $XS_c$ plotted against $X/X_K$.  The slope is a measure of the thermodynamic fragility, which is seen to decrease with increasing softness of interactions.}
\label{fig:therofrag}
\end{figure}

\subsection{Adam-Gibbs relation}
{We find that the kinetic fragility defined using the variable $X$ shows a trend that is opposite to that of the thermodynamic fragility, as interactions become softer}. To understand this disparity, we examine the Adam-Gibbs relation, which describes the dependence of the relaxation time on the configurational entropy. Further, assuming the validity of the  AG and VFT relations and $T^{(X)}_{k}=T^{(X)}_{VFT}$, we can infer the relationship between kinetic and thermodynamic fragility ({i.e.} $K^{(X)}_{VFT} = K_T^{(X)}/A$). To understand the role of the activation energy $A$, in explaining the opposite trends observed for the kinetic and thermodynamic fragilities, we study the AG relation for the diffusion coefficient $D_{A}$ and relaxation time $\tau$. The Fig. \ref{fig:AGplot} we show the Adam-Gibbs plots for the diffusion coefficient $D_{A}$ and relaxation time $\tau$, which illustrates that the activation energy $A$ is different for each model and its value decreases as the particle interaction become softer.  The relevant fit parameters are shown in Table \ref{tableX5}. The thermodynamic fragility estimated with the Adam-Gibbs relation, designated as Adam-Gibbs fragility ($K_{AG} = K_T/A)$ is listed in Table \ref{table1}. The AG fragility shows  agreement with the trend of the kinetic fragilities, although the AG fragility value are off by a factor of roughly two. The explanation can be seen in the plots in Fig. \ref{fig:AGplot} (c-d), which show the Adam-Gibbs plots with the x-axis $A/XS_{c}$. If the Adam-Gibbs relation is valid ideally, we expect the data to fall on straight lines of unit slope, with possible shifts along the y-axis owing to different possible values of the limiting values $D_0$ or $\tau_0$ of the diffusion coefficients and relaxation times. We find this not to be the case, with deviations from AG behaviour at low densities. Indeed the limiting values $D_0$ or $\tau_0$ obtained from Adam-Gibbs plots are different from those obtained from VFT fits (see Table \ref{tableX5}). Thus, from the above analysis, we conclude that (a) the Adam-Gibbs activation energies $A$ that go into defining the AG fragilities are non-trivially different for the different models studied, and (b) the AG relation itself is not well satisfied, pointing to possible deficiencies in the density-temperature scaling we perform. These aspects merit further investigation.

\begin{table}[h]
\centering
 \begin{tabular}{||c c c c||} 
 \hline
   & Model & $\tau_o$  or $1/D_o$   &  $A^{(X)}$  \\ [0.5ex] 
 \hline \hline	
$\tau$  &12,11	& 0.01897	& 1.07652   \\
&12,6	& 0.03885	&0.70002	\\	
&8,5	&  0.04698	&0.593871\\ 
$1/D_A$ &12,11	&  3.54671	&0.869835   \\
&12,6	& 7.79958	&0.551894	\\	
&8,5	&  6.30580	&0.4700  \\ [1ex] 
 \hline	          
\end{tabular}
\caption{Fit parameters for the Adam-Gibbs relation for scaled $\tau$, $D_A$ vs $1/XS_c$.}
\label{tableX5}
\end{table}

\begin{figure}[h!]
\centering
\includegraphics[scale=0.285]{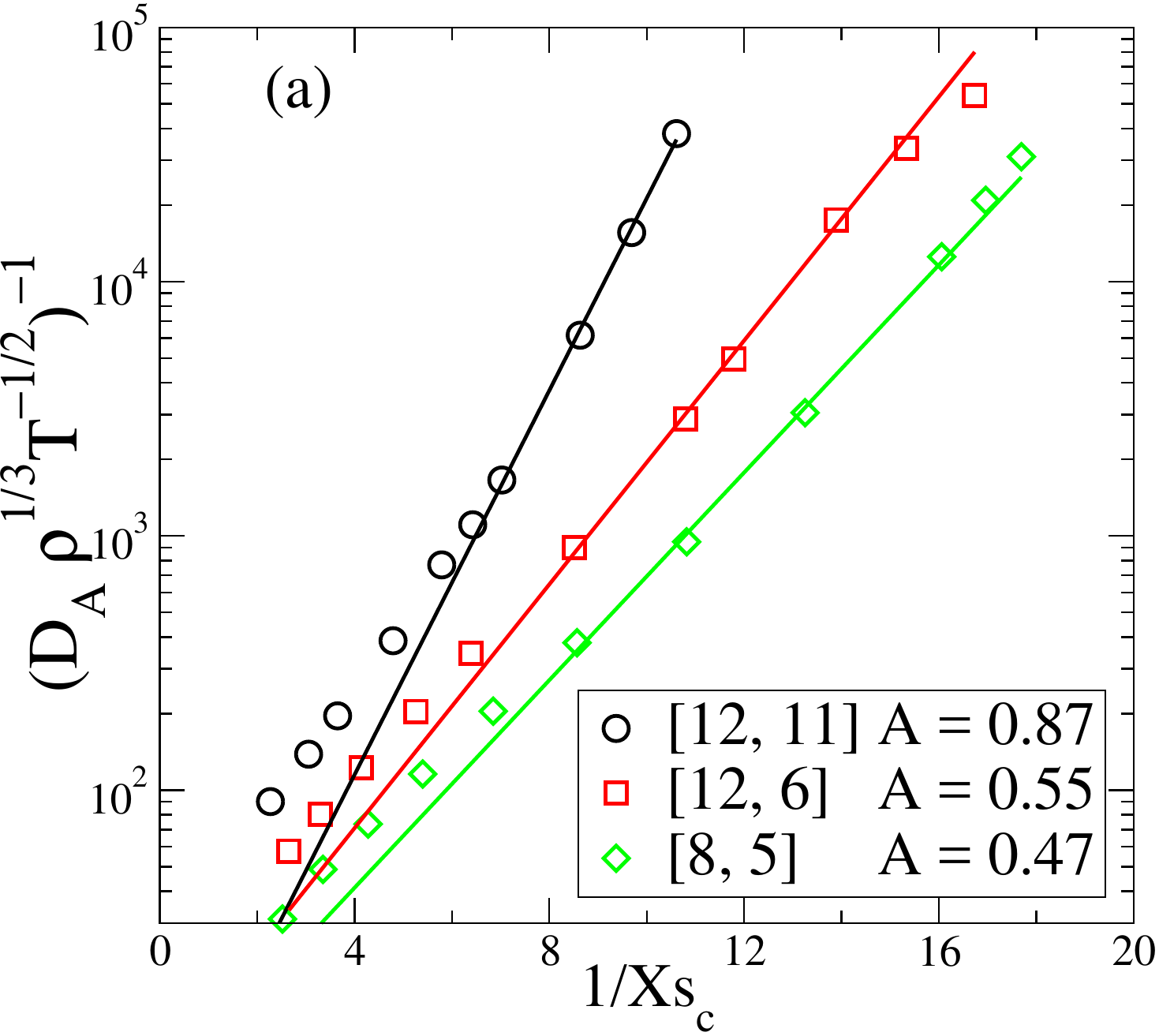}  \hspace{-2mm}
\includegraphics[scale=0.285]{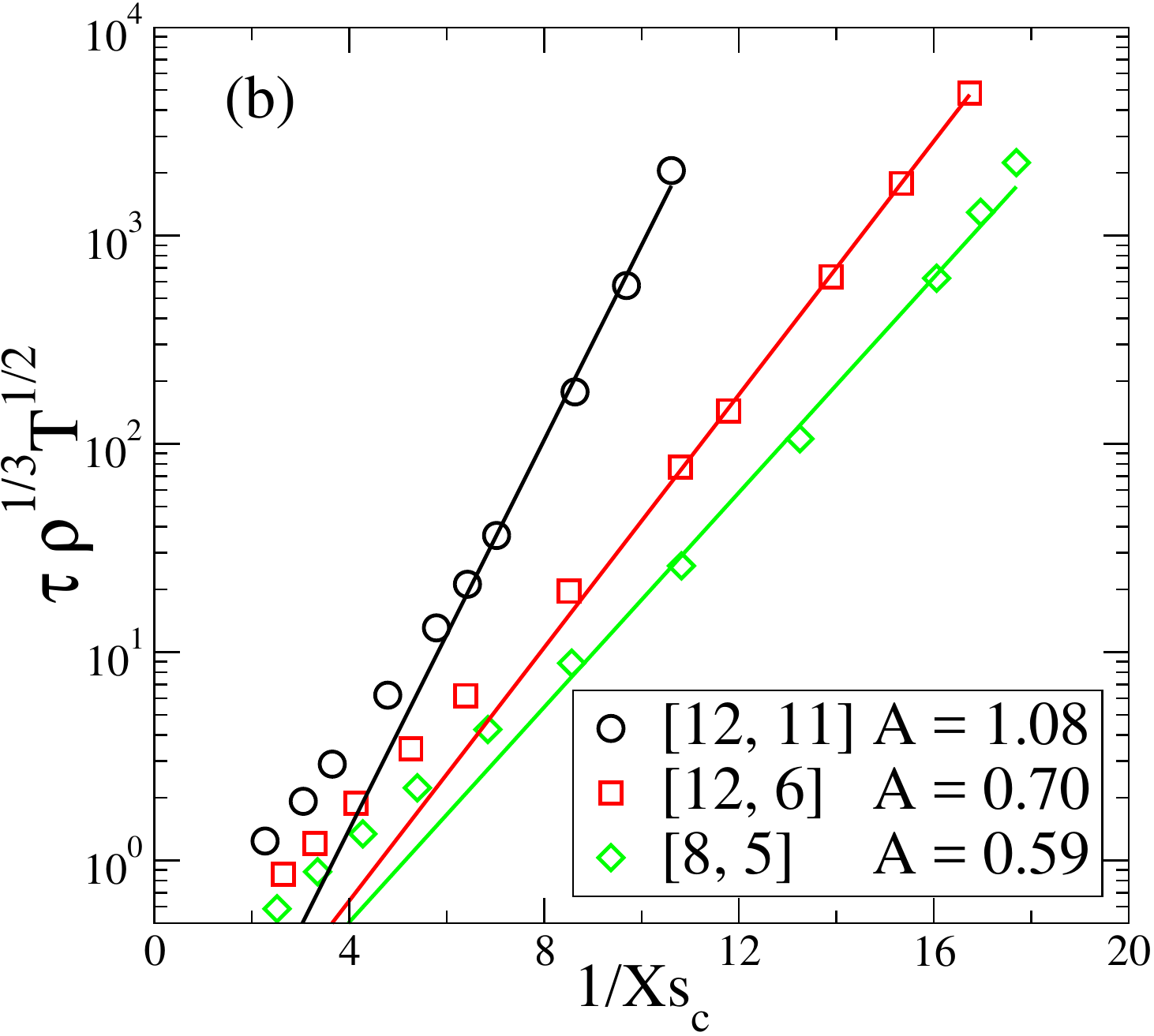}    \hspace{-2mm}
\includegraphics[scale=0.285]{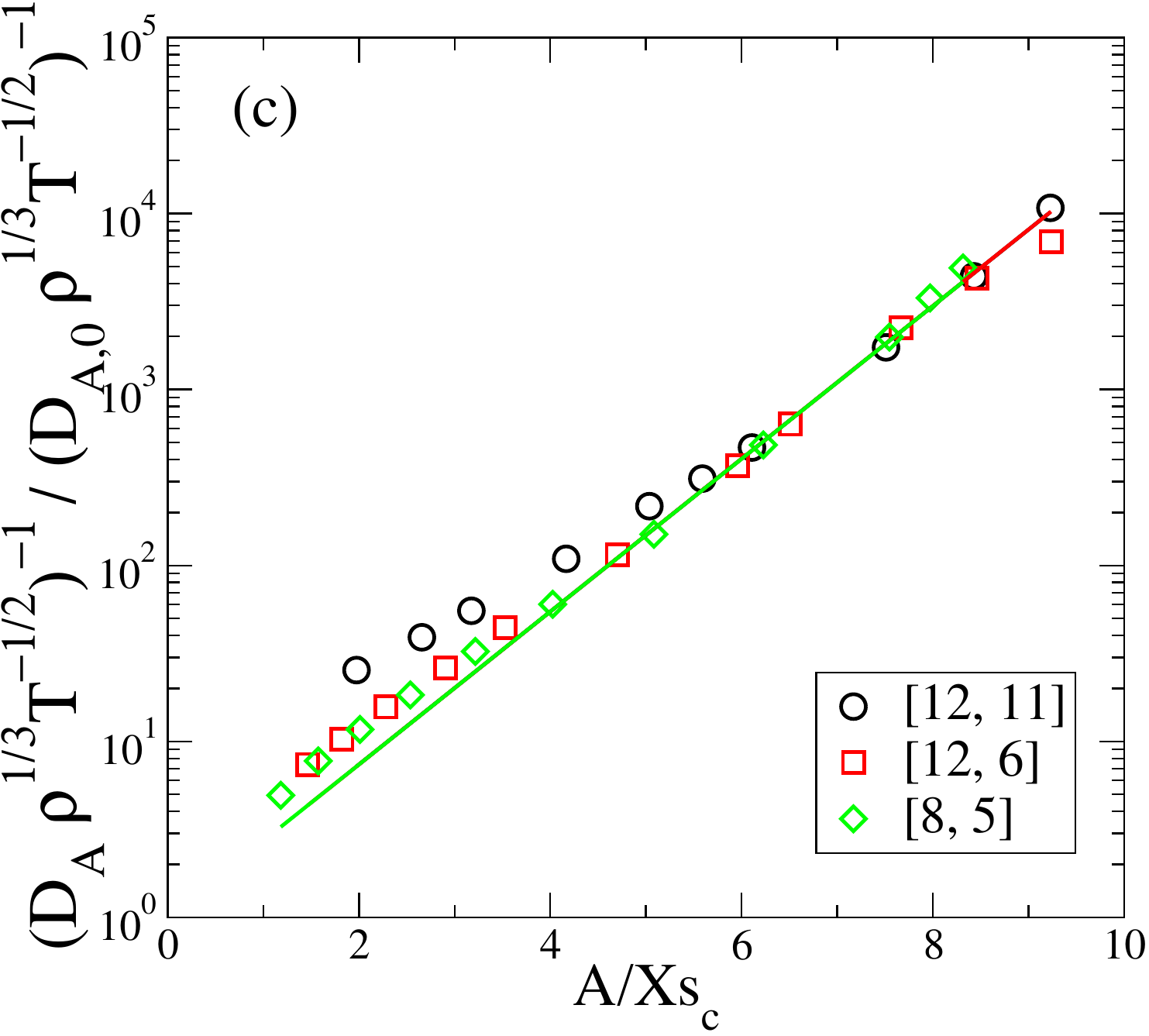}   \hspace{-2mm}
\includegraphics[scale=0.285]{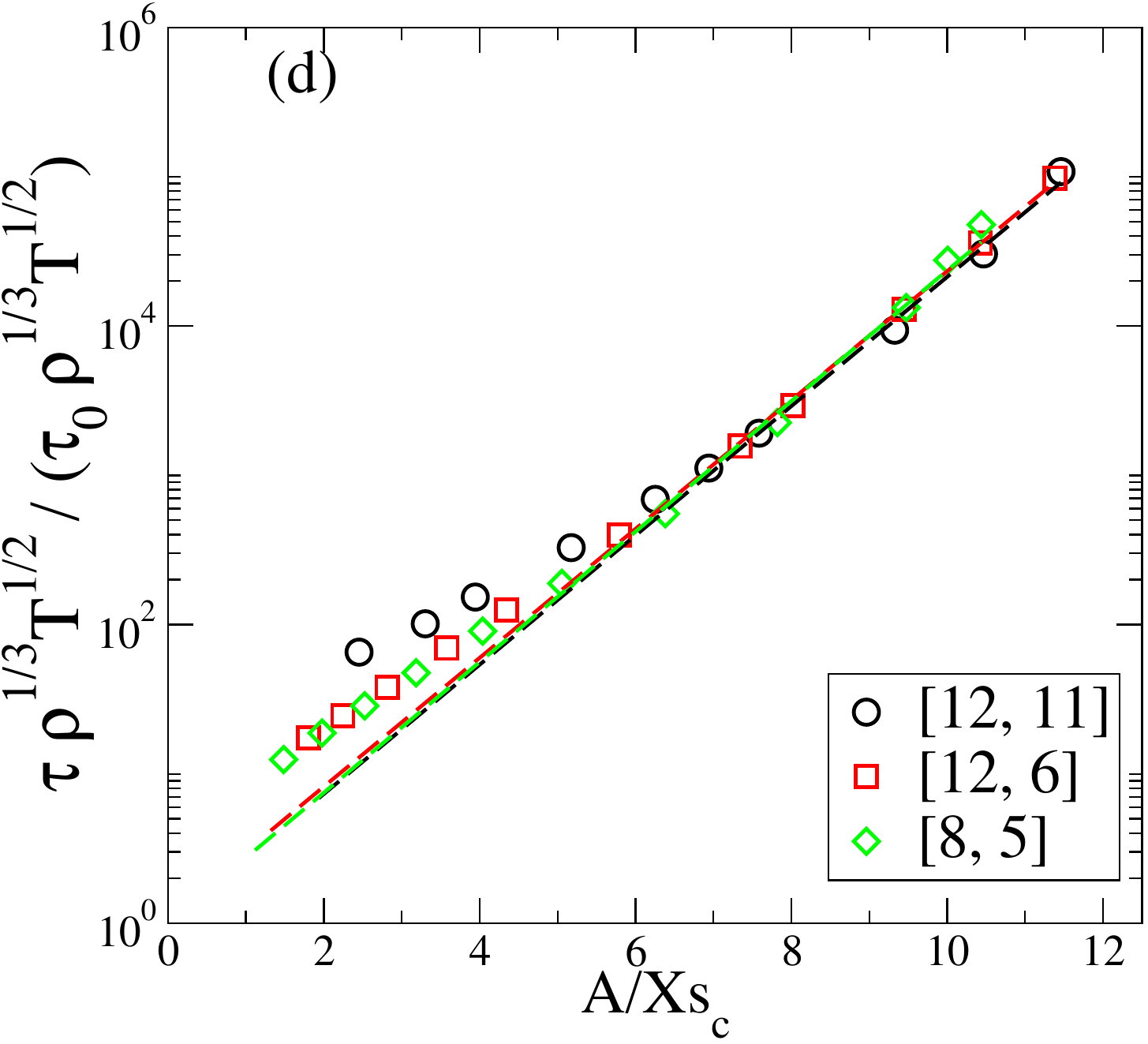}  \hspace{-2mm}
\caption{The Adam-Gibbs plots for (a) the diffusion coefficient and (b) relaxation time, for the models studied. The activations energy $A$ decreases as interactions becomes softer. The scaled Adam-Gibbs plot for (c) the diffusion coefficient and (d) relaxation time, which shows that the AG relation does not describe the data well for low densities.}
\label{fig:AGplot}
\end{figure}

\begin{table}[]
\centering
 \begin{tabular}{||c c c c c c c||} 
 \hline
  & Model& $K_{VFT}^{(\rho)}$ &  $K_{VFT}^{(X)}$ & $K_T^{(X)}$ & $A$ & $K_{AG}$\\ [0.5ex] 
 \hline \hline
 $\tau$        & 12,11  & 1.97 & 0.24 & 0.110 & 1.08 & 0.102\\ 
                & 12,6   & 1.48 & 0.25 & 0.096 & 0.70 & 0.137\\
                & 8,5    & 1.18 & 0.30 & 0.083 & 0.59 & 0.141\\
  $D_A$           & 12,11  & 1.94 & 0.25 & 0.110 & 0.87 & 0.126\\
                & 12,6   & 1.35 & 0.28 & 0.096 & 0.55 & 0.175\\
                & 8,5    & 1.14 & 0.30 & 0.083 & 0.47 & 0.177\\ [1ex] 
 \hline
 \end{tabular}
 \caption{The comparison of fragility parameters for various models.
 The kinetic fragility $K_{VFT}^{(\rho)}$ as function of softness shows that softer interactions correspond to stronger glass formers. However, the  $K_{VFT}^{X}$, defined from the VFT relation for $X$ displays the opposite behaviour. Further, the Adam-Gibbs fragility ($K_{AG}$), estimated from the thermodynamic fragility ($K_T^{(X)}$) and activation energy ($A$), displays consistent behaviour with $K_{VFT}^{X}$.}
\label{table1}
\end{table}

As a tentative step in probing the limitations of the density-temperature scaling, we perform analysis along the above lines, but using instead $\gamma$ values obtained empirically, from fitting $S_c$ data, by plotting  $TS_cln(D/D_0)$ against $\rho$, as shown in Fig. \ref{DT:AGgamma1} (a).  Fig. \ref{DT:AGgamma} (a) shows the plot of $X^{'}S_c$ vs $X^{'}/X^{'}_K$, and the corresponding thermodynamic fragilities are also shown in Table. \ref{table2}. {Fig.} \ref{DT:AGgamma} (b) 
shows the Angell plot of diffusion coefficients.  The corresponding kinetic fragilities for diffusion coefficients, labeled $K_{VFT}^{(X')}$ are indicated  in Table.
\ref{table2}.  Finally, Fig. \ref{DT:AGgamma} (c) shows the AG plots including the AG activation energies (listed in Table \ref{table2}) estimated from AG plots {\it vs.} $1/X^{'}S_{c}$, which indicate that the dynamical data now conform considerably better to the AG behaviour. Correspondingly, the AG fragilities, shown in Table \ref{table2}, agree much more closely with the kinetic fragilities.  This procedure, although more satisfactory than the one preceding, is based on an {\it ad hoc} estimate of the $\gamma$ exponents, and must thus be viewed as tentative. A more satisfactory approach may be to use as the scaled variable a more general function of the form $e(\rho)/T$ as has been suggested \cite{tarjus2,gnan2009,gnan2011}, which should be pursued in future work. 

\begin{table}[h]
\centering
 \begin{tabular}{||c c c c c c||} 
 \hline
&Model		&$K_{VFT}^{(X')}$ &	$K_T^{(X')} $	&	$A^{(X^{'})}$  & $K'_{AG}$ \\ [0.5ex]
\hline \hline
&12,11		 &0.197	   &0.077	&0.568	&0.135\\
&12,6        &0.206    &0.098   &0.560	&0.175\\
&8,5         &0.230    &0.046   &0.252	&0.181\\
\hline
 \end{tabular}
 \caption{Comparison of the fragilities, obtained using the scaling exponents $\gamma$ estimated empirically from the 
density dependence of the $TS_c ln(D/D_o)$,  as shown in Fig. \ref{DT:AGgamma1}. The kinetic and Adam-Gibbs fragilities show consistent trends with the softness of interactions, and agree better quantitatively than the results summarised in Table \ref{table1}.}
\label{table2}
\end{table}

\begin{figure}[h!]
\centering
\includegraphics[scale=0.28]{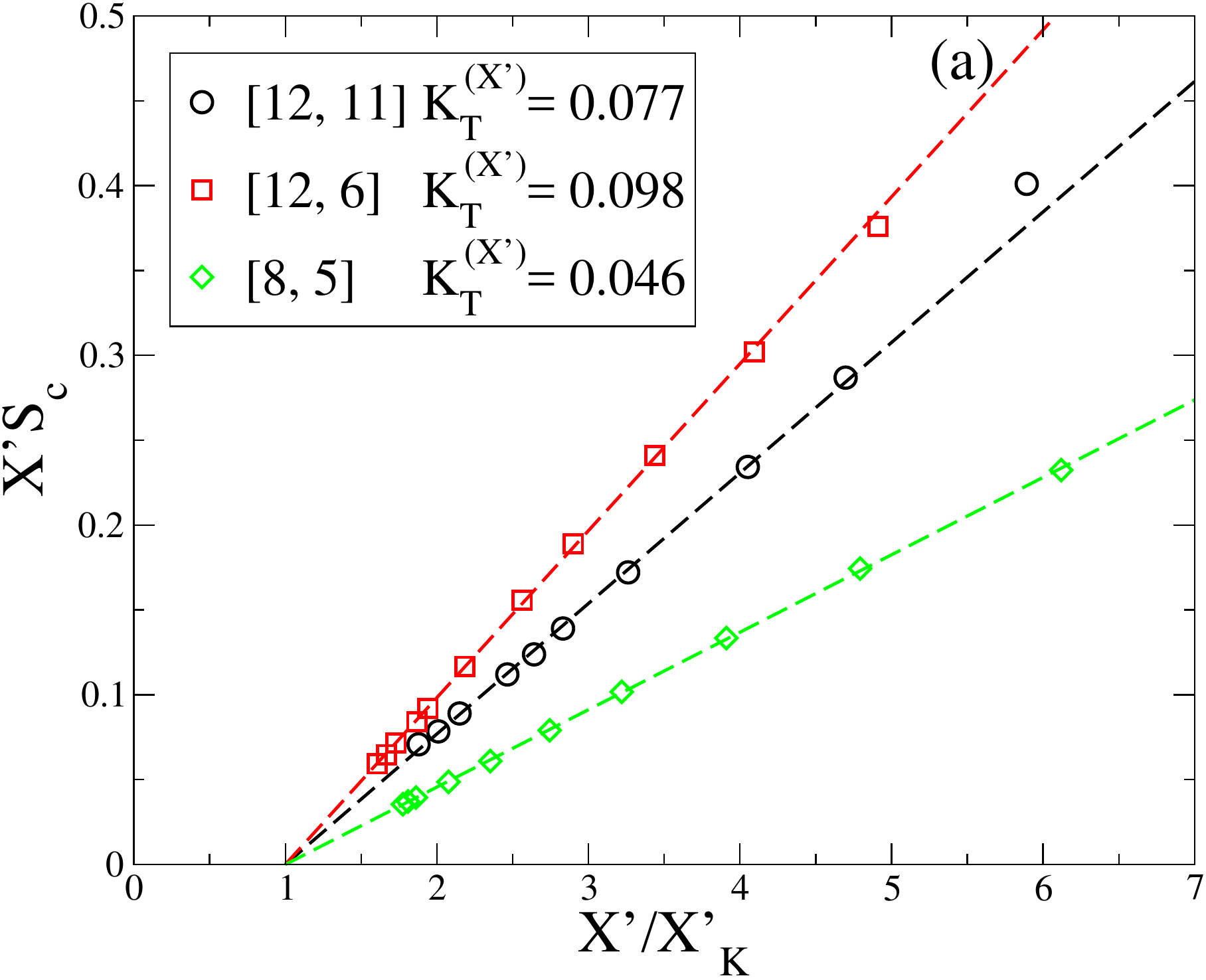}
\includegraphics[scale=0.28]{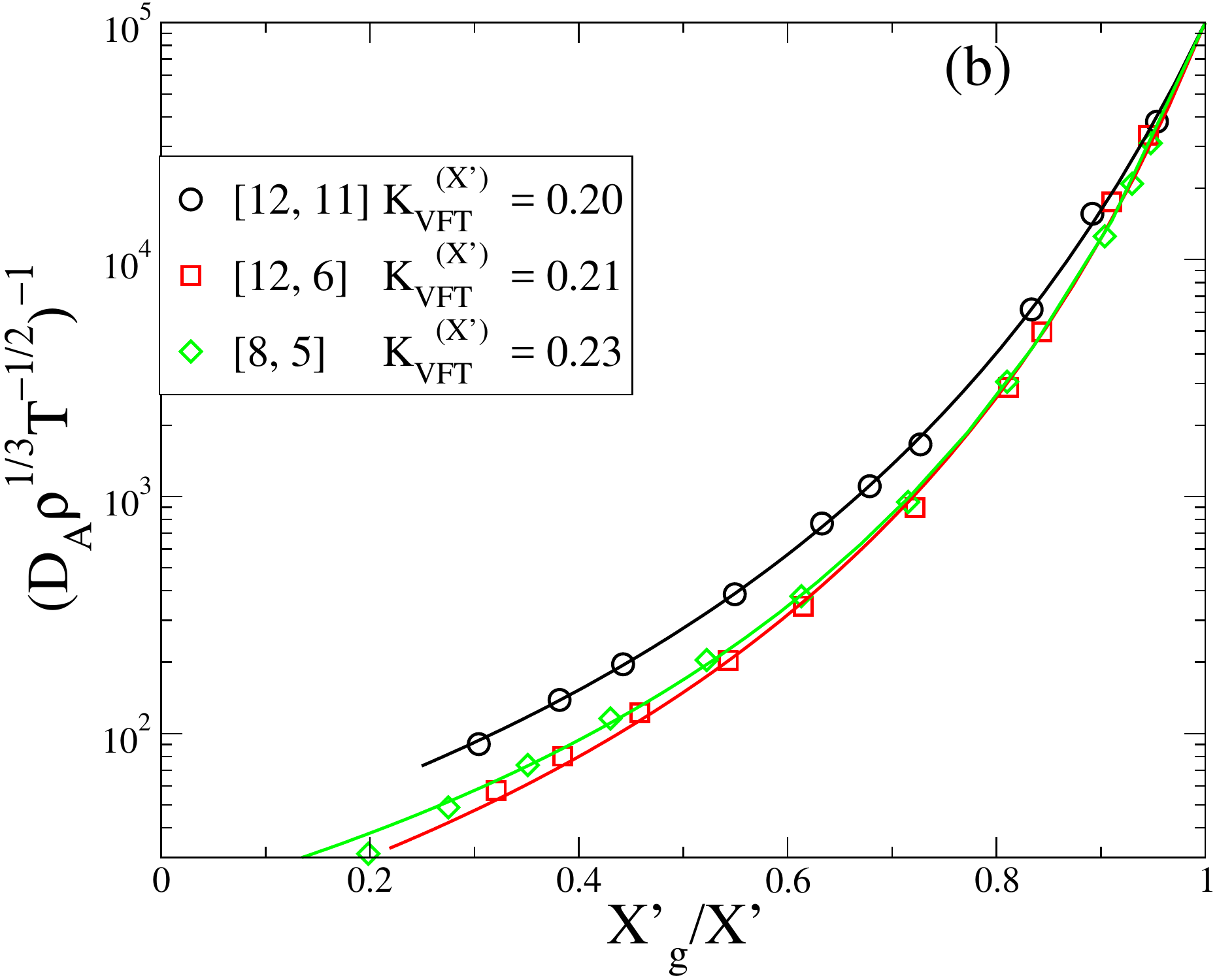} 
\includegraphics[scale=0.28]{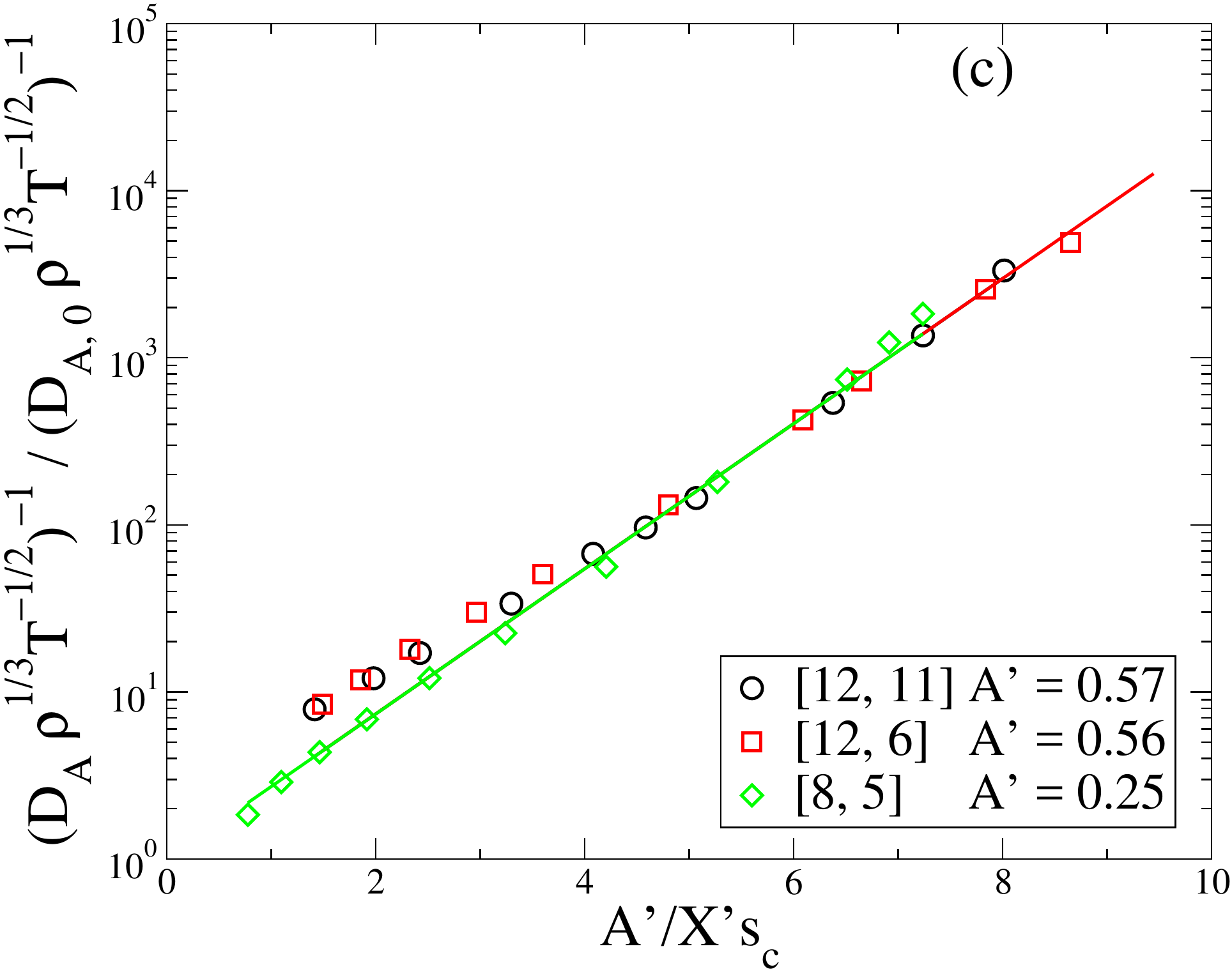}  
\caption{ (a) Thermodynamic fragility, (b) kinetic fragility  and (c) the Adam-Gibbs relation using  the scaling exponents $\gamma$ estimated empirically from the density dependence of the $TS_c ln(D/D_o)$,  as shown in Fig. \ref{DT:AGgamma1}. }
\label{DT:AGgamma}
\end{figure}

\section{Summary and Conclusions}

In this paper, we have attempted to address the question of how the fragility of glass formers depends on the softness of interactions. In particular, we have attempted to rationalise experimental observations that soft interactions lead to strong glass formers, and conflicting simulation results and analysis that suggest that soft interactions lead to more fragile glass formers. The key to rationalising these conflicting results appears to be in employing  a suitable variable to describe the changes in dynamics, and we have employed density-temperature scaling to attempt to do so. By employing a scaled variable involving temperature and density as the control parameter, we show that even when density is the physical parameter varies, softer interactions correspond to more fragile behaviour. To explain the behaviour by thermodynamic means, we compute the thermodynamic and Adam-Gibbs fragilities. We show that the thermodynamic fragility is smaller for softer interactions, consistent with previous results. Likewise, we show that the Adam-Gibbs activation energy plays a key role in determining kinetic fragilities. Unfortunately, we find that this parameter varies in a non-trivial way across the different models we study, and at present we do not have an appealing way of explaining or calculating this parameter. While the computation of the Adam-Gibbs fragility satisfactorily predicts the trend in fragility we observe from dynamical data, the quantitative accuracy is not very satisfactory. The limitations in this regard may arise either by the limited applicability of density-temperature scaling of the form we have employed in this work, or from the calculation of configurational entropy from a harmonic approximation, which may not be satisfied at low densities. Both these approximations should be visited in future work to have a better understanding of the variation of fragility in the series of systems we have studied.

\section*{Acknowledgement}
This paper is dedicated to the memory of Prof. Charusita Chakravarty, a respected and creative colleague, with a shared interest in understanding complex dynamics in liquids through thermodynamic approaches, whose untimely passing away has cut short a career that produced many insights and no doubt would have  generated many more. We seek inspiration in her restless spirit of wishing to delve deeper, and are saddened by the untimely end to her endeavors. 


\end{document}